\documentclass[11pt, onecolumn]{IEEEtran} 
\pdfoutput=1
\usepackage{amsmath}
\usepackage{amssymb}
\usepackage{amsfonts}
\usepackage{epsfig}
\usepackage{subfigure}
\usepackage{calc}
\usepackage{color}
\usepackage[all]{xy}
\usepackage{bm}
\usepackage{algorithmicx}
\usepackage{algpseudocode}
\usepackage{enumerate}
\usepackage{hyperref}

\newtheorem{defn}{Definition}
\newtheorem{thm}{Theorem}[section]
\newtheorem{cor}[thm]{Corollary}
\newtheorem{prop}{Proposition}

\newtheorem{lem}[thm]{Lemma}
\newtheorem{conj}[thm]{Conjecture}
\newtheorem{constr}[thm]{Construction}
\newtheorem{note}{Remark}

\newcommand{\bit}{\begin{itemize}}
\newcommand{\eit}{\end{itemize}}
\newcommand{\bcor}{\begin{cor}}
\newcommand{\ecor}{\end{cor}}
\newcommand{\beq}{\begin{equation}}
\newcommand{\eeq}{\end{equation}}
\newcommand{\beqn}{\begin{equation*}}
\newcommand{\eeqn}{\end{equation*}}
\newcommand{\bea}{\begin{eqnarray}}
\newcommand{\eea}{\end{eqnarray}}
\newcommand{\bean}{\begin{eqnarray*}}
\newcommand{\eean}{\end{eqnarray*}}
\newcommand{\ben}{\begin{enumerate}}
\newcommand{\een}{\end{enumerate}}
\newcommand{\bdefn}{\begin{defn}}
\newcommand{\edefn}{\end{defn}}
\newcommand{\bnote}{\begin{note}}
\newcommand{\enote}{\end{note}}
\newcommand{\bprop}{\begin{prop}}
\newcommand{\eprop}{\end{prop}}
\newcommand{\blem}{\begin{lem}}
\newcommand{\elem}{\end{lem}}
\newcommand{\bthm}{\begin{thm}}
\newcommand{\ethm}{\end{thm}}
\newcommand{\bconj}{\begin{conj}}
\newcommand{\econj}{\end{conj}}
\newcommand{\bconstr}{\begin{constr}}
\newcommand{\econstr}{\end{constr}}
\newcommand{\bpf}{\begin{proof}}
\newcommand{\epf}{\end{proof}}

\newcommand{\cmds}{\mbox{${\cal C}_{\text{\tiny MDS }}$}}
\newcommand{\gmds}{\mbox{$G_{\text{\tiny MDS}}$}}

\newcommand{\ccan}{\mbox{${\cal C}_{\text{can }}$}}

\newcommand{\clrc}{\mbox{${\cal C}_{\text{lrc }}$}}

\begin{document}

\title{ High-Rate Regenerating Codes Through Layering }
 \author{Birenjith Sasidharan and P. Vijay Kumar
\thanks{Birenjith Sasidharan and P. Vijay Kumar are with the Department of ECE, Indian Institute of Science, Bangalore 560 012, India (email: \{biren,vijay\}@ece.iisc.ernet.in).  This research is supported in part by the National Science Foundation under Grant 0964507 and in part by the NetApp Faculty Fellowship program.} }

\date{\today}
\maketitle

\begin{abstract}
In this paper, we provide explicit constructions for a class of exact-repair regenerating codes that possess a layered structure. These regenerating codes correspond to interior points on the storage-repair-bandwidth tradeoff, and compare very well in comparison to scheme that employs space-sharing between MSR and MBR codes. For the parameter set $(n,k,d=k)$ with $n < 2k-1$, we construct a class of codes with an auxiliary parameter $w$, referred to as canonical codes. With $w$ in the range $n-k < w < k$, these codes operate in the region between the MSR point and the MBR point, and perform significantly better than the space-sharing line. They only require a field size greater than $w+n-k$. For the case of $(n,n-1,n-1)$, canonical codes can also be shown to achieve an interior point on the line-segment joining the MSR point and the next point of slope-discontinuity on the storage-repair-bandwidth tradeoff. Thus we establish the existence of exact-repair codes on a point other than the MSR and the MBR point on the storage-repair-bandwidth tradeoff. We also construct layered regenerating codes for general parameter set $(n,k<d,k)$, which we refer to as non-canonical codes. These codes also perform significantly better than the space-sharing line, though they require a significantly higher field size. All the codes constructed in this paper are high-rate, can repair multiple node-failures and do not require any computation at the helper nodes. We also construct optimal codes with locality in which the local codes are layered regenerating codes. 

\end{abstract}

\section{Introduction} \label{sec:intro}

\subsection{Regenerating Codes} \label{sec:intro_reg}

In a distributed storage system, information pertaining to a single file is distributed across multiple nodes. In the present context, a file is a collection of $K$ symbols drawn from a finite field $\mathbb{F}_q$ of size $q$. Thus a file can be represented as a $(1\times K)$ vector over $\mathbb{F}_q$. A data collector should be able to retrieve the entire file by downloading data from any arbitrary set of $k$ nodes. Since the nodes are prone to failure, the system should be able to repair a failed node by downloading data from the remaining active nodes. In the framework of regenerating codes introduced in \cite{DimGodWuWaiRam}, a codeword is a $\mathbb{F}_q$-matrix of size $(\alpha \times n)$, where each column corresponds to the data stored by a single node.  A failed node is regenerated by downloading $\beta \leq \alpha$ symbols from any arbitrary set of $d$ nodes. These $d$ nodes are referred to as helper nodes. Since the entire file can be recovered from any arbitrary set of $k$ nodes, we must have

\bean
k \leq d \leq n-1.
\eean

The total bandwidth consumed for the repair of a single node equals $d\beta$ and is termed the repair bandwidth. Thus a regenerating code is parameterized by the ordered set, $\left(\mathbb{F}_q, (n,k,d), (\alpha, \beta), K\right)$. 

In the framework of regenerating codes, it is not required that the replacement node contain exactly the same symbols as did the failed node.  It is only required that following regeneration, the network possess the same properties with regard to data collection and node repair as it did prior to node failure.  Thus one distinguishes between functional and exact repair in a regeneration code, ~\cite{DimGodWuWaiRam}, \cite{WuDim}, \cite{RasShaKumRam_allerton}.     The present paper is concerned only with exact repair.   

A regenerating code is said to be linear if the encoded block of $(\alpha \times n)$ matrix is a linear transformation of the $(1\times B)$-size file vector.   Linear codes offer the advantage that data recovery and node regeneration can be accomplished through low-complexity, linear operations over the field $\mathbb{F}_q$.   The regenerating codes constructed in the present paper have the additional feature that no computations are needed at a helper node, a simple transfer of the contents of the helper node suffice.  We will term such regenerating codes as {\em help-by-transfer} regenerating codes.  This is distinct from the class of help-by-transfer regenerating codes discussed in \cite{ShaRasKumRam_rbt} which have the additional feature that no computations are needed even at the replacement node.

\subsection{The Classical Storage-Repair-Bandwidth Tradeoff} \label{sec:intro_tradeoff}

A major result in the field of regenerating codes is the proof in ~\cite{DimGodWuWaiRam} that uses the cut-set bound of network coding to establish that the parameters of a regenerating code must necessarily satisfy the inequality
\bea
K \leq \sum_{i=0}^{k-1} \min(\alpha, (d-i)\beta) .  \label{eq:tradeoff}
\eea
Optimal regenerating codes are those for which equality holds in \eqref{eq:tradeoff}.   It turns out that for a given value of $K,k,d$, there are multiple pairs $(\alpha,\beta)$ for which equality holds in \eqref{eq:tradeoff}.  It is desirable to minimize both $\alpha$ as well as $\beta$ since minimizing $\alpha$ reduces storage requirements, while minimizing $\beta$ results in a storage solution that minimizes repair bandwidth.  It is not possible to minimize both $\alpha$ and $\beta$ simultaneously and thus there is a tradeoff between choices of the parameters $\alpha$ and $\beta$.  The two extreme points in this tradeoff are termed the minimum storage regeneration (MSR) and minimum bandwidth regeneration (MBR) points respectively. The parameters $\alpha$ and $\beta$ for the MSR point on the tradeoff can be obtained by first minimizing $\alpha$ and then minimizing $\beta$ to obtain
\bea \label{eq:MSR_B} 
K &= & k \alpha \\ \label{eq:MSR_alpha} 
\alpha & = & (d-k+1)\beta 
\eea
Reversing the order leads to the MBR point which thus corresponds to
\bea \label{eq:MBR_B} 
K &= & dk - {k \choose 2}  \\ \label{eq:MBR_alpha} 
\alpha & = & d \beta .
\eea

\begin{figure}[ht]
\begin{minipage}[b]{0.45\linewidth}
\centering
\includegraphics[width=3.3in]{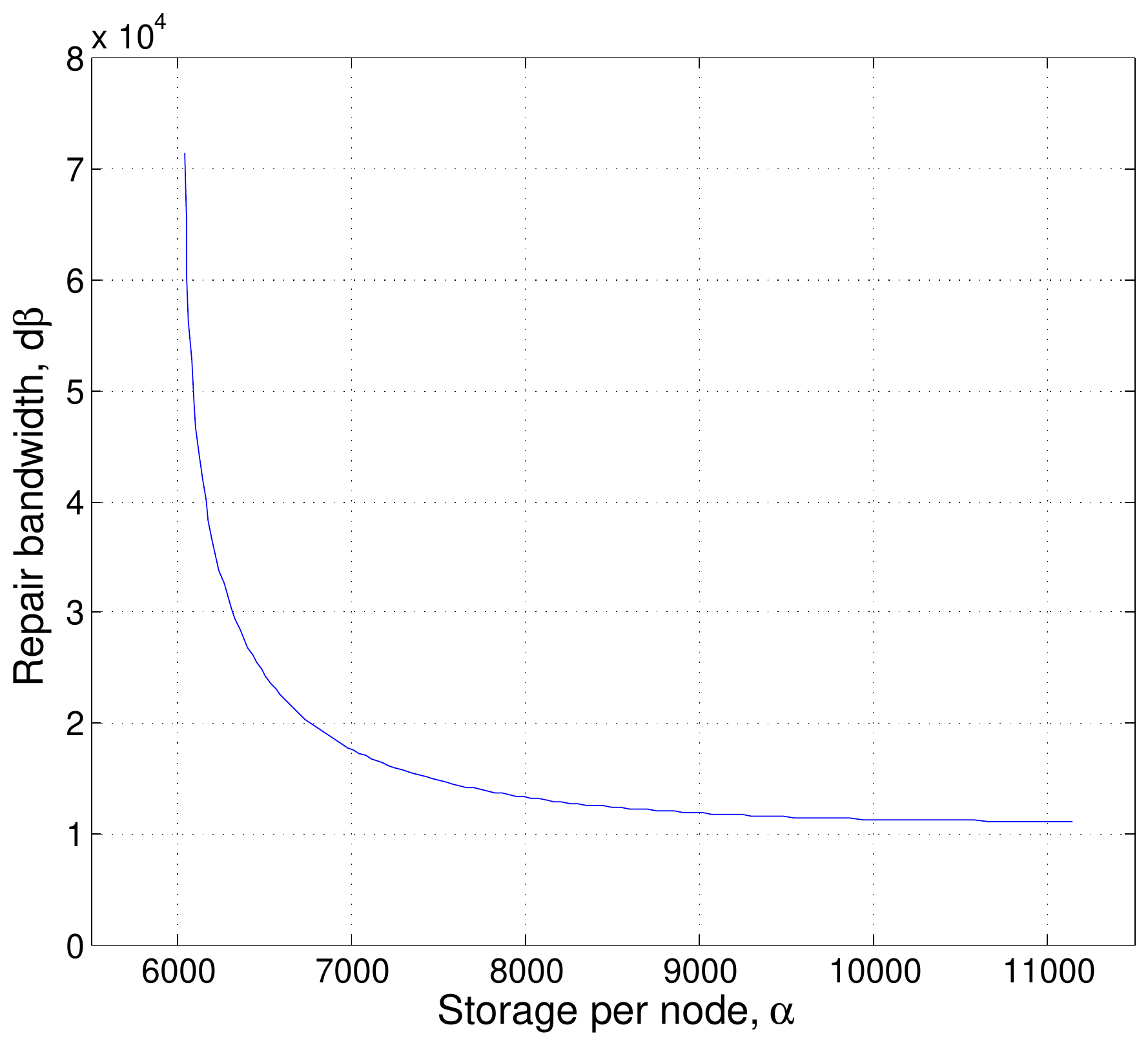}
\caption{Storage-repair-bandwidth tradeoff.  Here $[n=131, k=120, d=129, B=725360]$.}
\label{fig:tradeoff}
\end{minipage}
\hspace{0.5cm}
\begin{minipage}[b]{0.45\linewidth}
\centering
\includegraphics[width=3.3in]{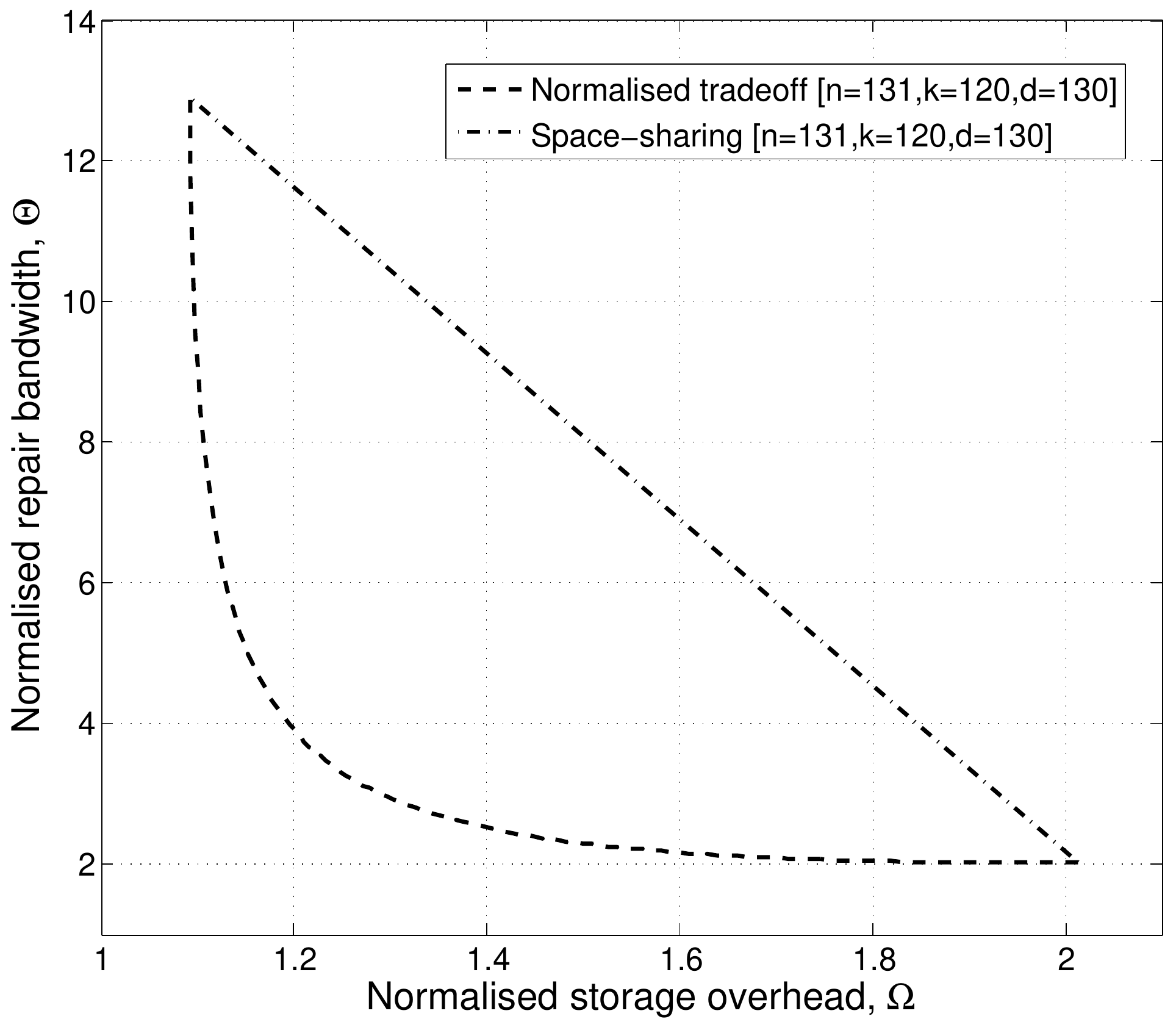}
\caption{Normalised storage-repair-bandwidth tradeoff.$[n=131, k=120, d=129]$}
\label{fig:normalised_tradeoff}
\end{minipage}
\end{figure}

The remaining points on the tradeoff will be referred to as interior points. As the tradeoff is a piecewise linear relation, there are $k$ points of slope discontinuity, corresponding to 
\bean
\alpha = (d-p)\beta, \ \ p \in \{0, 1, \cdots k-1 \}.
\eean
Setting $p=(k-1)$ and $0$ respectively yields the MSR and MBR points respectively. Thus the remaining values of $p \in \{1, \cdots k-2\}$ correspond to interior points. The tradeoff between $\alpha$ and $d\beta$ is plotted in Fig.~\ref{fig:tradeoff} for $(n=131,k=120,d=130)$ and filesize $K=725360$.

In \cite{ShaRasKumRam_rbt}, the authors proved that the interior points of the storage-repair-bandwidth-tradeoff cannot be achieved, under exact repair. This raises an open question as to how close one can come to the tradeoff at an interior point.  

\subsection{Normalised Storage-Repair-Bandwidth Tradeoff} \label{sec:intro_normalised}

In this subsection, we draw upon \cite{KamPraLalKum} and \cite{KamKum} to introduce a normalized version of the classical storage-repair-bandwidth tradeoff which we motivate as follows. Consider a situation where a user desires to store a file of size $K$ across $n$ nodes for a time period $T$ with each node storing $\alpha$ symbols. We follow~\cite{KamPraLalKum} and assume a Poisson-process model of node failures under which  the number of failures in time $T$ is proportional to the product of $T$ and the number of nodes $n$.  We also assume that there is cost associated with both node storage as well as with repair bandwidth. The cost of storage is assumed to be proportional to the amount of data stored, i.e., to $n\alpha$. The cost of a single node-repair is taken as the amount of data download to repair a node, i.e., $d\beta$. For simplicity, we only consider the case of single-node repairs in performance comparison, although a similar analysis can be carried out for the case of multiple node failures.  With this, it follows that if $\gamma(K,T)$ denotes the cost incurred to store a file of size $K$ for a time period $T$ using a particular coding scheme, then
\bea
 \gamma(K,T) &=& \left( \gamma_B nd\beta + \gamma_S n\alpha \right)T
\eea
for some proportionality constants $\gamma_B, \gamma_S$. Hence the average cost incurred in storing one symbol for one unit of time is given by
\bea
 \frac{\gamma(K,T)}{KT} &=& \gamma_B \frac{nd\beta}{K} + \gamma_S \frac{n\alpha}{K}.
\eea
We will refer to the quantities $\Omega := \frac{n\alpha}{K}$ and $\Theta := \frac{nd\beta}{K}$, as the storage overhead and normalized repair bandwidth of the code respectively. Thus the average cost is a linear combination of the normalized repair bandwidth $\Theta$ as well as the storage overhead $\Omega$.  The rate $R$ of a code is the inverse of $\Omega$, i.e., 
\bean
R & = & \frac{1}{\Omega}. 
\eean
When we set $d=n-\gamma$, and $\alpha = (d-p)\beta, \ p \in \{0,1,\cdots,k-1\}$, the tradeoff in \eqref{eq:tradeoff} translates to 
\bea
\Omega & \geq & \left( \frac{k}{n} - \frac{(k-p)(k-p-1)}{2n(n-\gamma)} \right)^{-1} = \ \ \Omega^{*} \label{eq:omegabound}\\
\Theta & \geq & \frac{n-\gamma}{n-\gamma-p}\left( \frac{k}{n} - \frac{(k-p)(k-p-1)}{2n(n-\gamma)} \right)^{-1} \ = \ \Theta^{*}
\eea
where $\Omega^{*}$ and $\Theta^{*}$ represent the minimum possible values of $\Omega$ and $\Theta$ respectively.

A plot of of $\Theta^{*}$ as a function of $\Omega^{*}$ when $\frac{p}{n}$ is varied, is referred to as the normalised tradeoff. Unlike in the classical tradeoff, points in the modified tradeoff do not correspond to a fixed file size $K$, neither to a fixed parameter set $[n,k,d]$.  The normalized tradeoff is parameterised for $\frac{k}{n}$ and $\frac{d}{n}$ where the tradeoff corresponds to the varying paramter $\frac{p}{n}$, which takes rational values bounded within $0$ and $\frac{k}{n}$. However, the plot shown in Fig.~\ref{fig:normalised_tradeoff} is for a fixed parameter set $[n=131,k=120,d=129]$. 

As in \cite{KamKum}, an asymptotic analysis of normalised storage-repair-bandwidth tradeoff can be done as $n$ scales to infinity. In this approach, the following quantities 
\bean
\kappa & = & \lim_{n \rightarrow \infty} \frac{k}{n} \\
\theta & = & \lim_{n \rightarrow \infty} \frac{p}{k}, \ \kappa \neq 0 \\
\Delta & = & \lim_{n \rightarrow \infty} \frac{d}{n} \\
\eean
are fixed as $n$ scales. If one assumes that $d=n-\gamma$, where $\gamma$ does not scale with $n$, we obtain $\Delta = 1$. Note that $\kappa \in [0 \ 1]$ and, $\theta \in [0 \ 1]$. Here, $\theta=0$ correspond to the MBR point, $\theta=1$ correspond to the MSR point, and $\theta \in (0 \ 1)$ correspond to the interior points of the storage-repair-bandwidth tradeoff. In this setting, we obtain an asymptotic version of the normalised storage-repair-bandwidth tradeoff as given below.
\bea
\Omega_a & \geq & \left( \kappa - \frac{(1-\theta)^2}{2(1-\theta\kappa)} \kappa^2 \right)^{-1} = \ \ \Omega_a^{*} \label{eq:omegabound_asymptote}\\
\Theta_a & \geq & \frac{1}{1-\theta\kappa}\left( \kappa - \frac{(1-\theta)^2}{2(1-\theta\kappa)} \kappa^2 \right)^{-1} 
 \\
 & = & \Theta_a^{*} \ = \ \frac{1}{1-\theta\kappa}\Omega_a^{*}  \label{eq:zeta_asymptote}
\eea

\begin{figure}[h!]
\begin{center}
\includegraphics[width=3.5in]{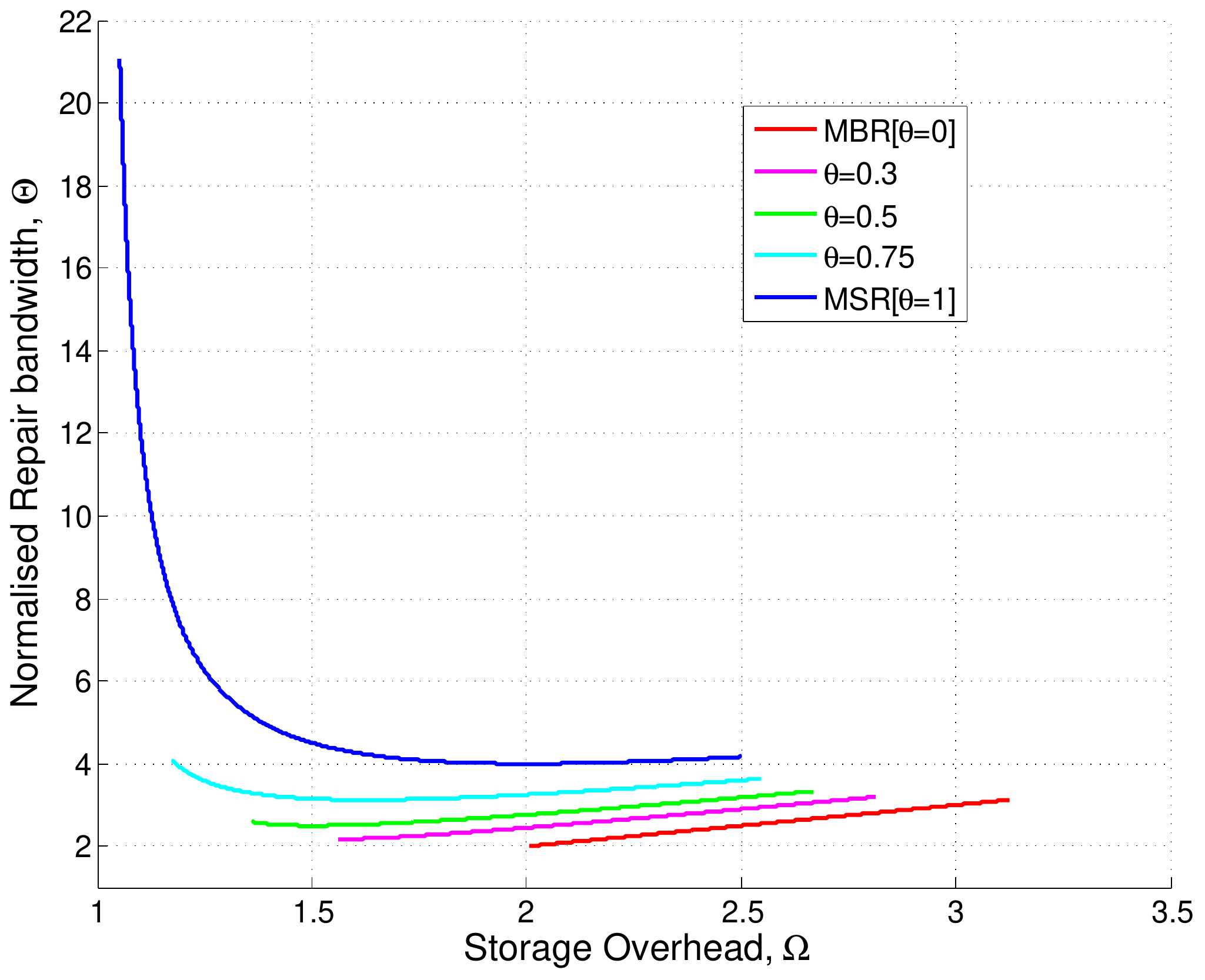}
\caption{Asymptotic normalised storage-repair-bandwidth tradeoff, as a function of $\kappa$, for various $\theta$.}
\label{fig:asymptotic_tradeoff}
\end{center}
\end{figure}

In \cite{KamKum}, the authors have used the asymptotic analysis to study the variation of the tradeoff with respect to $\kappa$, for various points of operation $\theta$. These plots, drawn in Fig.~\ref{fig:asymptotic_tradeoff}, showcase the importance of regenerating codes for the interior points of the tradeoff. From Fig.~\ref{fig:asymptotic_tradeoff}, it follows that for any fixed storage overhead,  repair bandwidth can be minimized by operating with the lowest value of $\theta$ that supports the given storage overhead. For example, if it is sufficient to build a distributed storage system with storage overhead $> 2$, then it is better to operate with $\theta=0$, i.e., MBR point. Similarly, operating at $\theta=1$, i.e. MSR point, is desirable only when required storage overhead is very close to $1$. When the permissible storage overhead falls in the range $1 < \Omega_a < 2$, it is desirable to use codes that operate in the range $0 < \theta < 1$, i.e. in the interior region of the tradeoff. 

\hspace*{2.0in}

\subsection{Existing Coding Schemes with Exact Repair} \label{sec:intro_existing}

Several coding schemes have been proposed in the literature in the exact-repair setting. In \cite{RasShaKum_pm}, a framework to construct exact-repair optimal regenerating codes at the MBR and MSR points is provided. The framework permits the construction of MBR codes for all values for $[n,k,d]$, and of MSR codes for $d \leq 2k-3$. In \cite{PapDimCad}, high-rate MSR codes with parameters $[n,k=n-2,d=n-1]$ are constructed using Hadamard designs. In \cite{TamWanBru}, high-rate MSR codes are constructed for $d=n-1$; here efficient node-repair is guaranteed only in the case of systematic nodes. A construction for MSR codes with $d=n-1\geq 2k-1$ is presented in  \cite{SuhRam} and \cite{ShaRasKumRam_ia}. The construction of MSR codes for arbitrary values of $[n,k,d]$ remains an open problem, although it has been proven in \cite{CadJafMalRamSuh} that exact-repair MSR codes exist for any parameter set $[n,k,d]$ as the filesize grows to infinity. In \cite{ShaRasKumRam_rbt}, a construction for a family of repair-by-transfer MBR codes is presented.  The construction of regenerating codes for a functional-repair setting may be found in \cite{HuShu} and \cite{ShaRasKumRam_ia}.  The nonexistence of exact-repair codes that achieve the classical storage-repair bandwidth tradeoff is proven in \cite{ShaRasKumRam_rbt}. 

\subsection{Vector Codes} \label{sec:intro_vector}

Regenerating codes can also be viewed as vector codes. An $[n, K, d_{\text{min}}, \alpha]$ linear \textit{vector code} ${\mathcal{C}}$  over a field $\mathbb{F}_q$ is a subset of $(\mathbb{F}_q^{\alpha})^n$ for some $\alpha>1$, such that given $\mathbf{c}, \mathbf{c}' \in \mathcal{C}$ and $a,b \in \mathbb{F}_q$, $a\mathbf{c} + b\mathbf{c}'$ also belongs to $\mathcal{C}$. A codeword of a vector code is a matrix in $\mathbb{F}_q^{\alpha \times n}$, and a code symbol of a codeword is a vector in $\mathbb{F}_q^{\alpha}$. As a vector space over $\mathbb{F}_q$, $\mathcal{C}$ has dimension $K$, termed the scalar dimension of the code. The Hamming distance between two codewords is the number of codesymbol vectors at which they differ. In this sense, the code has minimum distance $d_{\min}$. 

Associated with the vector code $\mathcal{C}$ is an $\mathbb{F}_q$-linear scalar code $\mathcal{C}^{(s)}$ of length $N= n\alpha$, where $\mathcal{C}^{(s)}$ is the collection of $(1\times n\alpha)$ vectors obtained by vectorising each codeword matrix in some prescribed order.   Given  a generator matrix $G$ for the scalar code ${\cal C}^{(s)}$, the first code symbol in the vector code is naturally associated with the first $\alpha$ columns of $G$ and so on.  We will refer to the collection of $\alpha$ columns of $G$ associated with the $i^{\text{th}}$ code symbol ${\bf c}_i$ as the $i^{\text{th}}$ thick column.   We will refer to the columns of $G$ themselves as thin columns in order to avoid confusion, and thus there are $\alpha$ thin columns per thick column of the generator matrix.

\subsection{Locality} \label{sec:intro_locality}

Codes for distributed storage have been studied from other perspectives, different from the setting of regenerating codes. One prominent direction is related to codes with locality~\cite{GopHuaSimYek}. In this class of codes, a failed node is repaired by downloading entire data from a few set of nodes. Thus the property of locality allows to minimise the number of node accesses during repair. If locality-property holds only for systematic nodes, then it is referred to as information locality, and if it holds for all nodes, it is referred to as all-symbol locality. Scalar codes(i.e., $\alpha = 1$) with locality was introduced in~\cite{GopHuaSimYek}, for the case of single symbol erasure, and subsequently extended in \cite{PraKamLalKum}, for the case of multiple erasures. An upperbound on the minimum distance of the scalar code with locality was derived in above papers. Scalar codes with information locality that are optimal with respect to the aforesaid bound were constructed in \cite{HuaCheLi}. Optimal scalar all-symbol local codes were constructed in \cite{PraKamLalKum} and \cite{HanMonAlf}. Another class of codes\cite{OggDat}, named as homomorphic self-repairing codes, constructed using linearized polynomials also turns out to be optimal scalar all-symbol codes. Recently, the concept of locality was studied for vector codes (i.e., $\alpha > 1$) in \cite{KamPraLalKum}, \cite{RawKoySilVis} and \cite{PapDim}, and thereby making this class of codes to be a comparable alternative to regenerating codes. Codes combining benefits of regenerating codes and codes with locality were constructed in \cite{KamPraLalKum} and \cite{RawKoySilVis}. In ~\cite{RawKoySilVis}, the authors consider codes with all-symbol locality where the local codes are regenerating codes. Bounds on minimum distance are provided and a construction for optimal codes with MSR all-symbol locality based on linearized polynomials (rank-distance codes) is presented.

\subsection{Gabidulin Codes} 

Let $\mathcal{G} = \{g(x)=\sum_{i=0}^{D-1}g_i x^{q^i}\mid g_i \in \mathbb{F}_q\}$ denote  the set of all linearized polynomials of $q$-degree $\leq (D-1)$ over $\mathbb{F}_{q^N}$, and let $\{P_i\}_{i=1}^K$, $N \geq K \geq D$, be a collection of linearly independent elements over $\mathbb{F}_q$ in $\mathbb{F}_{q^N}$. Consider for each $g\in {\cal G}$, the vector $(g(P_1),g(P_2),\cdots,g(P_K))$. By representing each element $g(P_i)$ as an $N$-element vector over $\mathbb{F}_q$, we obtain an $(N \times K)$ matrix over $\mathbb{F}_q$.  The resultant collection of $q^D$ matrices turns out to form a maximal rank distance (MRD) code known as the Gabidulin code~\cite{Gab}.  In the current paper, we will in several places deal with vectors of the form $(g(P_1),\cdots,g(P_K))$, and it follows that these may also be regarded as codewords drawn from the Gabidulin code. 

\subsection{Results}

In this paper, we first construct an $(n, k, d=k)$-regenerating code having a layered structure which we term as the canonical code \ccan.  This code has two auxiliary parameters $w$ and $\gamma$ satisfying $w \geq 2, \gamma \geq 1, w+\gamma\leq n$ and only requires field size $q > w+\gamma$.   We show how starting from a canonical code, it is possible to build a second class of layered regenerating codes with $k<d$ by making suitable use of linearized polynomial evaluations (or equivalently, codewords in the Gabidulin code)  as is done in \cite{RawKoySilVis}. These codes will be referred to as non-canonical regenerating codes. The extension to the case $k<d$ requires however, an expansion in field size from $q$ to $q^K$ where $K$ is the scalar dimension of the underlying canonical code. These codes allows help-by-transfer repair(``uncoded'' repair) and are of high-rate. 

We also show that the canonical code with $\gamma < w < k$ always perform better than space-sharing code. Recently, Chao et al. proposed a construction of exact-repair codes \cite{TiaAggVai} using Steiner systems that achieves points better than the space-sharing line. \footnote{Their paper appeared in the public literature only after the initial submission of our paper on arXiv.} They consider constructions for $d=n-1$.  For the particular case of $d=n-1$ and $k=n-2$, the performance of their code is identical to the construction in the present paper when our construction is specialized to the same parameter set $d=n-1, k=n-2$. 

Our constructions with $k=d=n-1$ achieve an interior point of the storage-repair-bandwidth tradeoff, that is in the middle of the MSR point and the next point of slope-discontinuity. Recently, Chao \cite{Tia} has characterized the optimal storage-repair-bandwidth tradeoff of $(4,3,3)$-exact repair codes. It turns out that the $(4,3,3)$-canonical code appearing in this paper also achieves the same optimal region.

Finally, we construct codes with local regeneration following the techniques in \cite{PraKamLalKum}, \cite{RawKoySilVis}, in which the local codes correspond to the canonical code.

\subsection{Performance of Codes} \label{sec:intro_performance}

\begin{figure}[h!]
\begin{minipage}[b]{0.45\linewidth}
\centering
\includegraphics[width=3.5in]{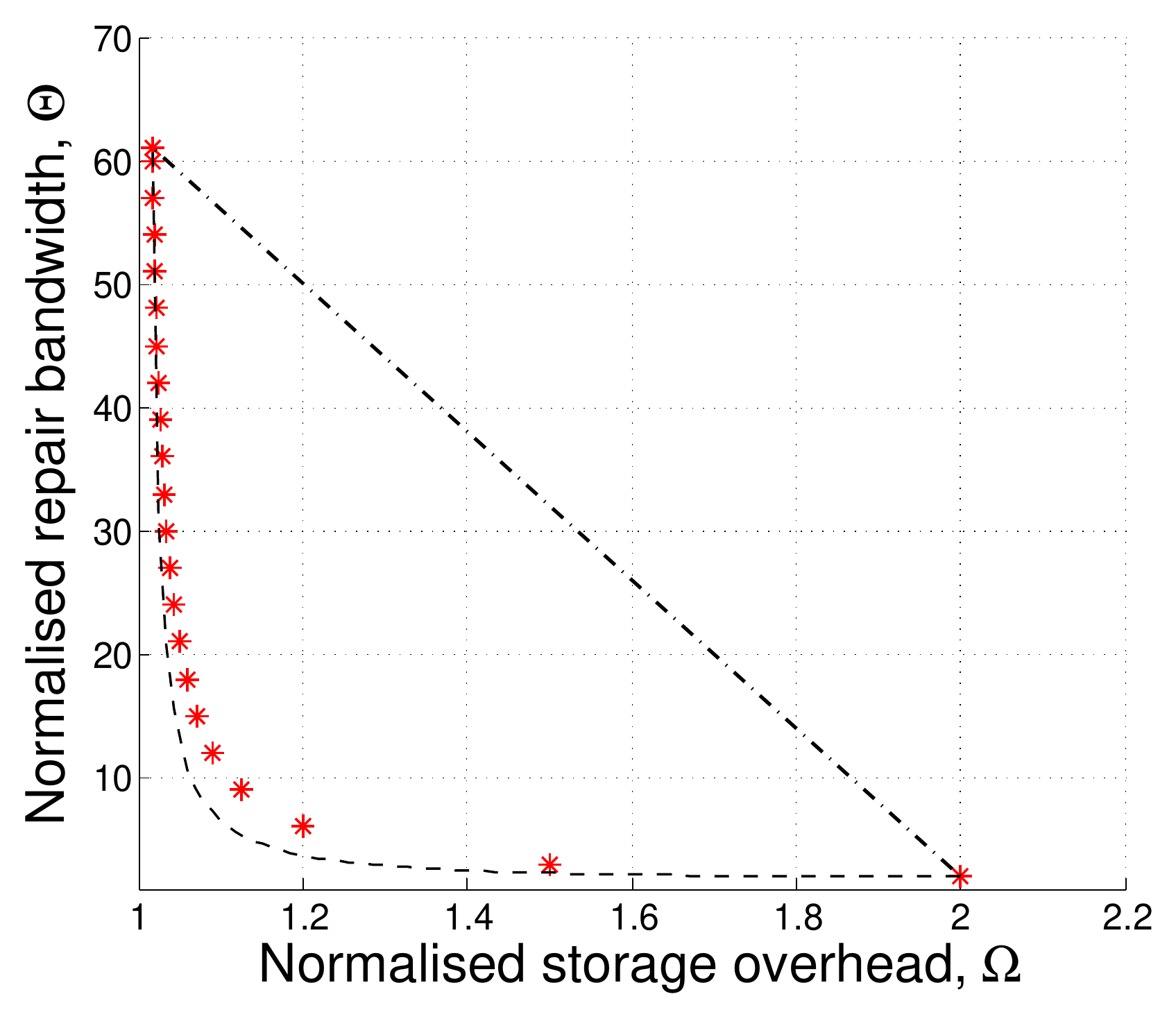}
\caption{Plot comparing the performance of the canonical code with $(n=61,k=d=60)$ for varying $w \in \{2,5,8,\ldots,59\}$. The MBR point (which is the degenerate case with $w=1$) and the MSR point are also marked. With $w=59$, an interior point on the tradeoff between the MSR point and the next point of slope-discontinuity is achieved.}
\label{fig:comparison_can1}
\end{minipage}
\hspace{0.5cm}
\begin{minipage}[b]{0.45\linewidth}
\centering
\includegraphics[width=3.5in]{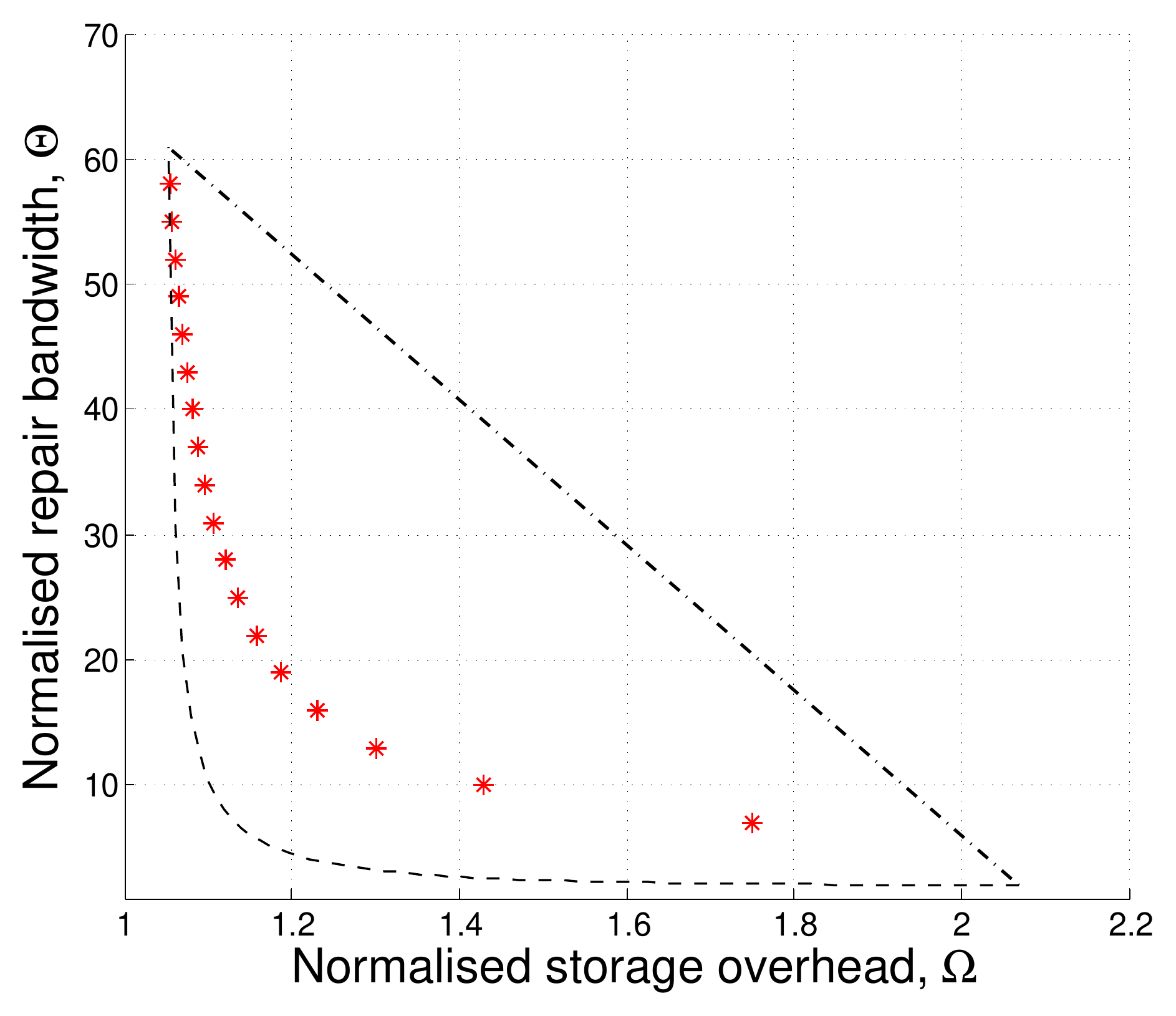}
\caption{Plot comparing performance of the canonical regenerating code with $(n=61,k=58,d=58)$ whiel varying $w \in \{4,7,10,\ldots,55\}$.}
\label{fig:comparison_can2}
\end{minipage}
\vspace{0.8cm}
\begin{minipage}[b]{0.45\linewidth}
\centering
\includegraphics[width=3.5in]{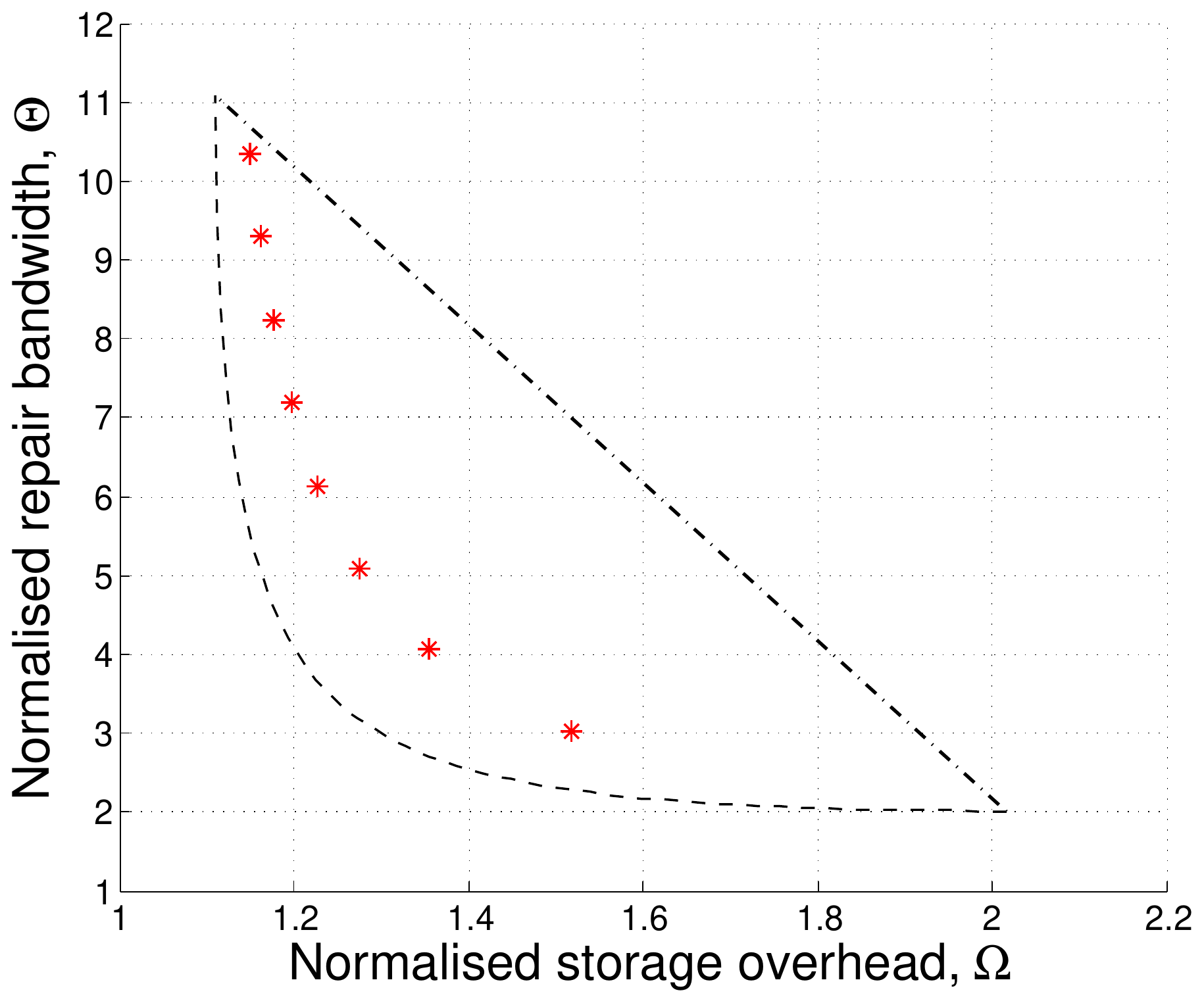}
\caption{Plot comparing performance of the non-canonical layered regenerating code with $(n=61,k=55,d=60)$ while varying $w \in \{2,3,\ldots,9\}$.}
\label{fig:comparison_lay}
\end{minipage}
\hspace{1cm}
\begin{minipage}[b]{0.45\linewidth}
\centering
\includegraphics[width=3.5in]{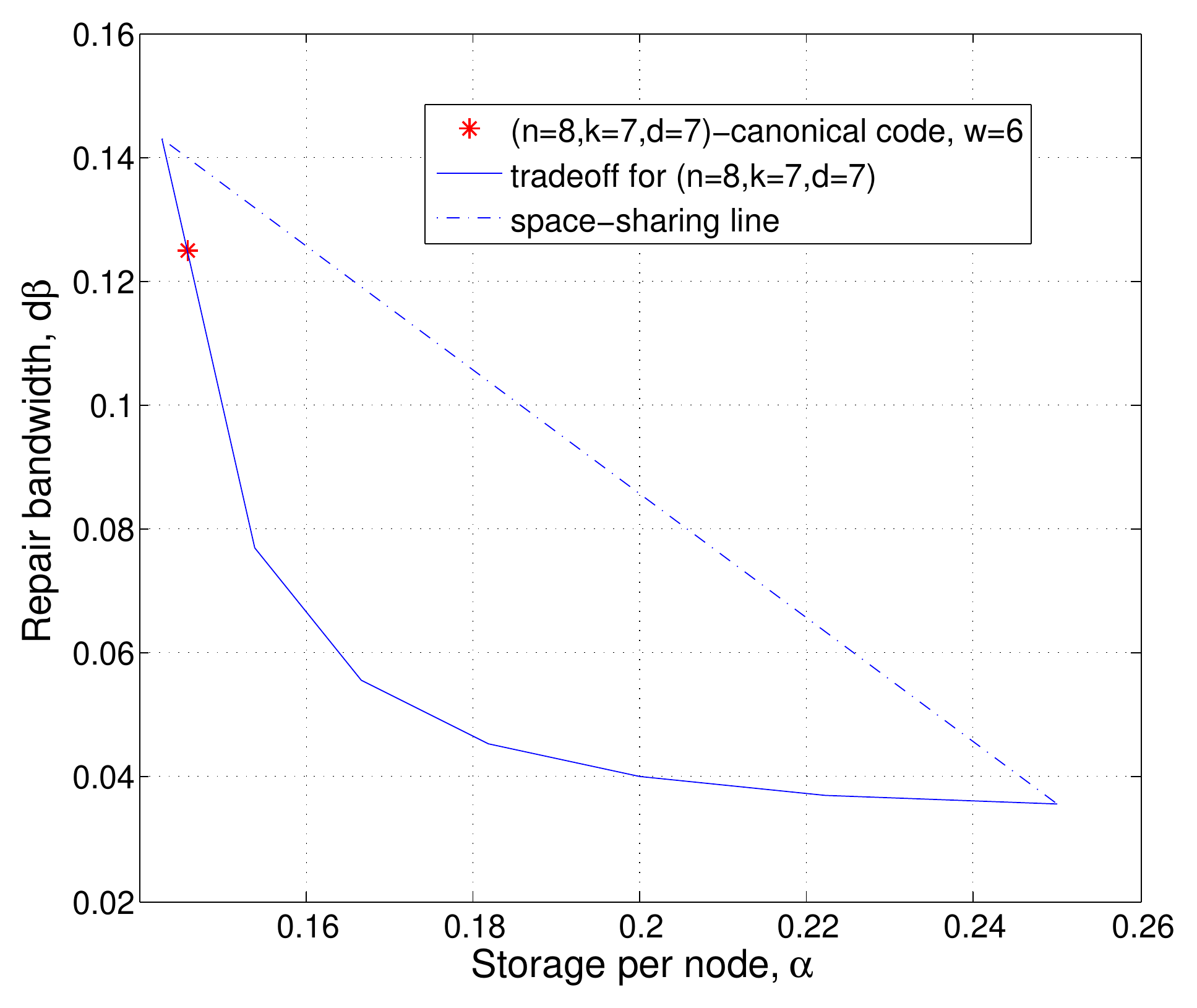}
\caption{$(8,7,7)$-canonical code with $w=6$ achieves the interior point.}
\label{fig:interior}
\end{minipage}
\end{figure}

The performance of this class of codes is compared against MBR and MSR codes using the normalized tradeoff. The layered codes operate in the interior region between the MSR and MBR points, and the auxiliary parameter $2 \leq w \leq k$ turns out to determine the specific interior point in the tradeoff. For a wide range of parameters $(n,k,d)$, these codes outperform codes that space-share between MSR and MBR codes~\footnote{Exact-repair MSR codes are not known to exist for every value of $(n,k,d)$-tuple. Hence the achievability of the space-sharing line joining MSR point and MBR point is not always guaranteed.}.  Figures \ref{fig:comparison_can1} and \ref{fig:comparison_can2} show the respective performance of canonical codes with $(n=61,k=60,d=60)$ and $(n=61,k=58,d=58)$. For the case of $(n=61,k=60,d=60)$, and interior point on the tradeoff between the MSR point and the next point of slope-discontinuity is achieved with $w=59$. Achievability of interior point by canonical construction is depicted in the classical storage-repair-bandwidth plot in Fig.~\ref{fig:interior} for the parameter set $(n=8,k=7,d=7)$ with auxiliary parameter $w=6$. The performance of non-canonical layered regenerating codes with $(n=61, k=55, d=60)$ is shown in Fig.\ref{fig:comparison_lay}.  As can be seen in plots, the codes come close to the tradeoff in terms of performance.

\section{Construction of the $(n,k,d=k)$-Canonical Layered Regenerating Code} \label{sec:can_coprime}

In this section, we will describe the construction of a family of high-rate, $((n,k,d=k), (\alpha,\beta),K_c)$ regenerating codes indexed by two  auxiliary parameters $w, \gamma$ satisfying $w \geq 2, \gamma \geq 1, w+\gamma\leq n$.  The code has a layered structure, and we will have $d=k$. The code will be simply referred to as canonical code. The construction we provide in this section, assumes $(n,w+\gamma)=1$.  The general case of $(n,w+\gamma)>1$ will be considered in the next section.  

\subsection{Construction of the Canonical Code $\mathcal{C}$} \label{sec:can_coprime_constrn} 

The construction will make use of certian other parameters derived from $w$ and $\gamma$ as defined below.
\bean 
L & = & \frac{1}{n}{n \choose w+\gamma} \ \ \ \text{ (number of patterns)} \\
V & = & \frac{1}{w} \text{lcm}(w,w+1,\cdots,w+\gamma-1)  \ \ \ \text{ (repetition factor (of each pattern))} \\\\
M & = & L V \ \ \ \text{ (number of layers)} \\
K_c & = & LVnw \ \ \ \text{ (scalar dimension of the canonical code)} .
\eean

\begin{figure}[h!]
\begin{center}
\includegraphics[scale=0.6]{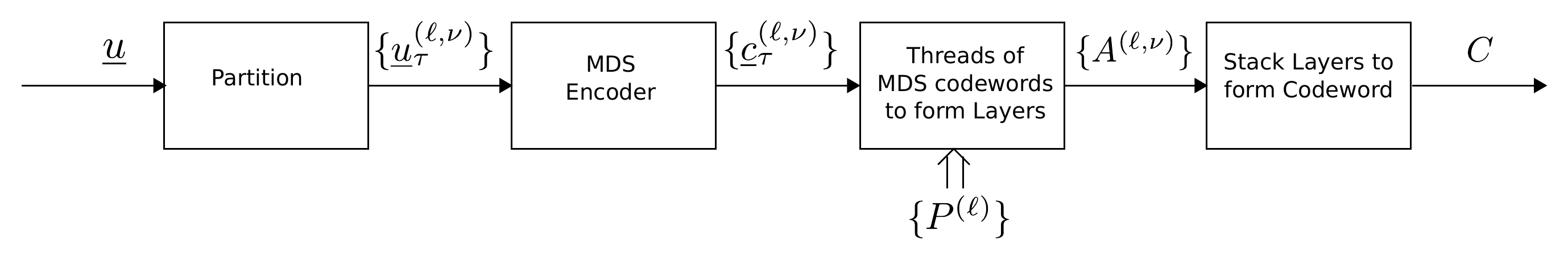}
\caption{Encoder of the canonical layered regenerating code.}
\label{fig:can_code_construction}
\end{center}
\end{figure}

The structure of the canonical code, can be inferred from Fig.~\ref{fig:can_code_construction} which shows the four-step process by which the incoming message vector $\underline{u}$ is encoded:
\ben[(a)]
\item The $K_c$-tuple message vector $\underline{u}$ is first partitioned into $LVn$ $w$-tuples: 
\bean
\underline{u} \in \mathbb{F}_q^{K_c} & \Rightarrow & \left\{ \underline{u}_{\tau}^{(\ell,\nu)} \in \mathbb{F}_q^w \mid 1 \leq \ell \leq L, \ \ 1 \leq \nu \leq V, \ \ 0 \leq \tau \leq n-1 \right\}. 
\eean
\item Each $w$-tuple is then encoded using an $[w+\gamma,w,\gamma+1]$ MDS code to yield $LVn$ codewords
\bean
\underline{u}_{\tau}^{(\ell,\nu)} \in \mathbb{F}_q^w & \Rightarrow & \underline{c}_{\tau}^{(\ell,\nu)} \in \mathbb{F}_q^{w+\gamma}.
\eean
\item The collection of $n$ codewords $\{ \underline{c}_{\tau}^{(\ell,\nu)} \}_{\tau=0}^{n-1} $ is then ``threaded'' to form a layer $A^{(\ell\,\nu)}$ of the code matrix:
\bean
\{ \underline{c}_{\tau}^{(\ell,\nu)} \}_{i=0}^{n-1}  & \Rightarrow & A^{(\ell,\nu)} .
\eean
This threading is carried out with the help of a pattern $\pi^{(\ell)}$.  The nature of a pattern and the threading process are explained below in Sections~\ref{sec:can_coprime_pattern},\ref{sec:can_coprime_threads}.
\item The $LV$ layers are then stacked to form the code matrix
\bean
C & = & \left[ \begin{array}{c} A^{(1,1)} \\ A^{(1,2)} \\ \vdots \\ A^{(L,V)} \end{array} \right].
\eean 
\een

\subsection{Patterns} \label{sec:can_coprime_pattern}

There are ${n \choose w+\gamma}$ subsets of $[n]$ that are of size $(w+\gamma)$. Let us partition these subsets into equivalence classes by declaring two elements to be equivalent if one is a cyclic shift of the other.  Given our assumption that $(n,w+\gamma)=1$, all equivalence classes will contain precisely $n$ elements and hence the number of equivalence classes is given by  $L=\frac{1}{n}{n\choose w+\gamma}$.   Let  
\bean
\left\{ \pi^{(\ell)} \mid 1 \leq \ell \leq L \right\} & = & \left\{ (\pi^{(\ell)}_1, \pi^{(\ell)}_2, \cdots, \pi^{(\ell)}_{w+\gamma} ) \mid  1 \leq \ell \leq L \right\}, 
\eean
be the collection of subsets obtained by selecting one subset from each equivalence class. We will assume that the elements within each of the subsets $\pi^{(\ell)}$ are 
ordered in ascending numerical order, i.e., 
\bean
\pi^{(\ell)}_1 < \pi^{(\ell)}_2 < \cdots < \pi^{(\ell)}_{w+\gamma}, \ \  \text{  for all  } \ell. 
\eean
We will associate with each such ordered subset, a collection of $n$ two-dimensional patterns, each of size $(w+\gamma)\times (w+\gamma)$.  This collection includes the fundamental pattern:
\bean
P^{(\ell)}(0) & = & \left\{ \left( i,\pi^{(\ell)}_i \right)  \mid 1 \leq i \leq w+\gamma \right\},
\eean
as well as its $n$ (columnar) cyclic shifts
\bean
P^{(\ell)}(\tau) & = &  \left\{ \left( i,\pi^{(\ell)}_i \oplus \tau \right)  \mid 1 \leq i \leq w+\gamma \right\}, \ \ 1 \leq \tau \leq (n-1) ,
\eean
in which $\pi^{(\ell)}_i \oplus \tau$ is addition modulo $n$.  Given a pattern $P^{(\ell)}(\tau)$ we will refer to the $(w+\gamma)$-tuple
\bean
\pi^{(\ell)} \oplus \tau & = &  (\pi^{(\ell)}_1 \oplus \tau, \ \pi^{(\ell)}_2 \oplus \tau, \cdots, \ \pi^{(\ell)}_{w+\gamma} \oplus \tau ), 
\eean
as its (columnar) footprint.   Thus the footprint of a fundamental pattern $P^{(\ell)}(0)$ is simply given by $\pi^{(\ell)}$. 

\subsection{Threading Codewords to Form a Layer} \label{sec:can_coprime_threads}

We fix $(\ell,\nu)$ and hence describe the threading process as it applies to the $(\ell,\nu)$th layer.  Consider the collection of $n$ codewords $\{\underline{c}_{\tau}^{(\ell,\nu)} \}_{\tau=0}^{n-1}$ associated to a layer.  The symbols of the $\tau$th codeword $\underline{c}_{\tau}^{(\ell,\nu)}$, $0 \leq \tau \leq n-1$, are placed (in any arbitrary order) into the $n$ locations \bean
P^{(\ell)}(\tau) & = &  \left\{ \left( i,\pi^{(\ell)}_i \oplus \tau \right)  \mid 1 \leq i \leq w+\gamma \right\}, \ \ 1 \leq \tau \leq (n-1) ,
\eean
identified by the pattern $P^{(\ell)}(\tau)$.  We might also refer to this codeword of this erasure code as a thread. The threading yields a $(w+\gamma \times n)$ matrix which we will denote by $A^{(\ell,\nu)}$.  The threading process is illustrated in Figure~\ref{fig:threading_process}. 
We then repeat this process for each layer, i.e., for all pairs $(\ell,\nu)$.  Finally we vertically stack the matrices $A^{(\ell,\nu)}$ to obtain the code matrix as described above. With this the encoding process is complete. 

\begin{figure}
\bean
\begin{array}{||c|c|c|c|c|c||} \hline \hline 
\text{ * }  &  \hspace{0.15in} & & & &   \\ \hline 
 &      &\text{ * }   &  & &   \\ \hline 
& &    &\text{ * }  & &   \\ \hline 
   & & &    &\text{ * }  &  \\ \hline 
& & & &  &\text{ * }  \\ \hline \hline 
  \end{array} & & \begin{array}{||c|c|c|c|c|c||} \hline \hline 
c_{11}  &  \hspace{0.15in} & & & &   \\ \hline 
 &      &c_{12}   &  & &   \\ \hline 
& &    &c_{13}   & &   \\ \hline 
   & & &    &c_{14}  &  \\ \hline 
& & & &  &c_{15}  \\ \hline \hline 
  \end{array} \\
\begin{array}{||c|c|c|c|c|c||} \hline \hline 
  & c_{21}  & \hspace{0.15in} & & &   \\ \hline 
 &      &   & c_{22} & &   \\ \hline 
& &    &   & c_{23} &   \\ \hline 
   & & &    &  & c_{24} \\ \hline 
c_{25} & & & &  &  \\ \hline \hline 
  \end{array}
& & 
\begin{array}{||c|c|c|c|c|c||} \hline \hline 
c_{11}  &  c_{21}    &c_{31}   & c_{41}   &c_{51}  &c_{61}  \\ \hline 
 c_{52}  &  c_{62}    &c_{12}   & c_{22}   &c_{32}  &c_{42}  \\ \hline 
  c_{43}  &  c_{53}    &c_{63}   & c_{13}   &c_{23}  &c_{33}  \\ \hline 
   c_{34}  &  c_{44}    &c_{54}   & c_{64}   &c_{14}  &c_{24}  \\ \hline 
    c_{25}  &  c_{35}    &c_{45}   & c_{55}   &c_{65}  &c_{15}  \\ \hline \hline 
  \end{array}.
\eean 
\caption{Illustrating the threading process.  The top left matrix uses an $*$ to identify the elements of the two-dimensional pattern $P^{(\ell)}(0)$.  The top right matrix shows the entries of a codeword $\underline{c}_0^{(\ell,\nu)}$ being inserted into the locations identified by the pattern.  The bottom left shows the codeword $\underline{c}_1^{(\ell,\nu)}$ inserted into the locations identified by $P^{(\ell)}(1)$.  The bottom right shows the completely filled in layer $A^{(\ell,\nu)}$.}
\label{fig:threading_process}
\end{figure}

\subsection{Parameters of the Canonical Code} \label{sec:can_coprim_parameters}

\subsubsection{Parameters $n,\alpha$}  \label{sec:can_coprime_nalpha}

The parameter $n$ is simply the block length of the code \ccan, viewed as a vector code with symbol alphabet $\mathbb{F}_q^{\alpha}$. The value of $\alpha$ can be computed from nature of the construction and is given by 
\bean
\alpha & = & LV(w+\gamma) \\
& = & \frac{1}{n}{n \choose w+\gamma} \frac{V_1}{w} (w+\gamma), \\
& = &  \frac{\text{lcm}(w,w+1,\cdots,w+\gamma-1) }{w}{n-1 \choose w+\gamma-1}.
\eean 

\subsubsection{Parameters $d,\beta$} \label{sec:can_coprime_dbeta}

We next note that the value of $d$ can be no less than $n-\gamma$ for otherwise, it would not be possible in some instances to repair a failed node.  This follows from the fact that the symbols of each MDS code are spread across $(w+\gamma)$ distinct nodes and that to repair a failed symbol in an $[w+\gamma,w,\gamma+1]$ MDS code, one needs access to at least $w$ symbols of the codeword.   Conversely, it follows that if $d=n-\gamma$, then every failed node can be repaired. We will set $d=n-\gamma$ here.  It remains to establish that repair of a failed node can be accomplished by connecting to $d$ nodes and downloading a {\em fixed} number $\beta$ of symbols from each of the $d$ helper nodes. 

It will be convenient in our analysis to assume that along with the given failed node (say node $\eta_1$), there are $\gamma-1$ other nodes (say, nodes $\eta_i$, $i=2,3,\ldots,\gamma$) that have also failed and that the remaining $d=n-\gamma$ nodes are acting as the helper nodes.  Let us assume further, that node $h$ is one of the helper nodes.  Our interest is in determining the number of symbols that need to be transferred from node $h$ to node $\eta_1$ for the purposes of node repair.    We had noted earlier in describing the construction of the canonical code, that each layer $A^{(\ell,\nu)}$ of the canonical code is composed of $n$ MDS codewords $\{\underline{c}_{\tau}^{(\ell,\nu)}\}_{\tau=0}^{n-1}$.  The codeword $\underline{c}_{\tau}^{(\ell,\nu)}$ is placed in the locations associated to the pattern $P^{(\ell)}(\tau)$.  We will refer to the $n$ MDS codes as threads in the description below. 

Node $h$ can transfer one symbol to the replacement for node $\eta_1$ iff there is a thread in some layer to which both nodes $\eta_1$ and $h$ contribute code symbols.  We now break up our count according to the total number $p$ of nodes that have now failed, but which previously contributed a symbol to the erasure code thread.  More specifically, we are counting the number of threads such that 
\bit
\item both nodes $\eta_1$ and $h$ contribute a single code symbol to that thread
\item $(p-1)$ of the nodes $\{\eta_i \mid 2 \leq i \leq \gamma\}$ each contribute one code symbol to the thread, the remaining failed nodes do not contribute any code symbol to the thread
\eit
The total number of such threads, across all the $L$ distinct layers in the code matrix is given by 
\bean
{\gamma-1 \choose p-1} {n-\gamma-1 \choose w+\gamma-p-1}.
\eean 
Within the erasure code, the situation is that $p$ symbols have been erased and thus a total of $w+\gamma-p$ symbols can serve as helper nodes for node $\eta_1$ of which node $h$ is one.  Since any $w$ nodes suffice to help node $\eta_1$ recover from the erasure, it suffices if node $h$ ``on average'' contributes a fraction 
\bean
\frac{w}{w+\gamma-p} 
\eean
of code symbols.    We can ensure that this average is realized by calling upon the $V$ repetitions of each layer.  The number $V$ has been chosen such that for all $p$, 
\bean
\frac{w+\gamma-p}{w} \mid V.
\eean
Thus we can ensure that the helper node will always pass on 
\bean
V \frac{w}{w+\gamma-p} 
\eean
code symbols when counted across all $V$ repetitions of the corresponding erasure code. It follows that the value of $\beta$ and $d$ are given by 
\bea \label{eq:beta_small_d}
\beta & = & V \ \sum_{p=1}^\gamma {\gamma-1 \choose p-1} {n-\gamma-1 \choose w+\gamma-p-1} \frac{w}{w+\gamma-p} \ , \\ 
d & = & n - \gamma . \label{eq:small_d}
\eea

As a check, we note that each column contains $\alpha=LV (w+\gamma)$ symbols, each of which requires the transfer of $w$ symbols to enable repair.   Since there are a total of $(n-\gamma)$ helper nodes, we must have that 
\bean
\beta (n-\gamma) & = & w \alpha ,
\eean
i.e., $\beta$ must equal 
\bea \nonumber 
\beta & = & \frac{w}{(n-\gamma)} \alpha \\ \label{eq:beta_2_small_d} 
& = & \frac{w}{(n-\gamma)} \frac{1}{n} {n \choose w+\gamma}  V(w+\gamma) .
\eea
It can be verified that the values for $\beta$ obtained in \eqref{eq:beta_small_d} and \eqref{eq:beta_2_small_d} are the same. 

\subsubsection{Determining $k, K$ and Code Rate $R$} \label{sec:can_coprime_k}

Arguing as above, if $k<(n-\gamma)$, we will fail to decode at least one thread. Hence $k \geq (n-\gamma)$.  On the other hand, by connecting to $d=(n-\gamma)$ we can recover the entire data and hence $k=d$.  The scalar dimension of the code is clearly given by $K_c=LVnw$.  Not surprisingly, the rate $R$ of the code is given by $\frac{w}{w+\gamma}$.  

\section{$(n,k,d=k)$-Canonical Code when $(n,w+\gamma)\neq 1$}\label{sec:can_gen}

We consider the general case when $(w+\gamma,n) \neq 1$ and let the integer $g$ be defined by setting 
\bean
\frac{n}{g} & = & (n,w+\gamma). 
\eean
The differences in the case of $(w+\gamma,n)\neq 1$ arise out of how patterns are identified in the canonical code. 

\subsection{Patterns} \label{sec:can_gen_patterns}

We partition as before, the ${n \choose w+\gamma}$ subsets of $[n]$ of size $(w+\gamma)$ into equivalence classes by declaring two subsets to be equivalent if one is a cyclic shift of the other.  This time, however, different equivalence classes will be of different size.    The number of elements in an equivalence class will always be of the form $gr$ with $r$ dividing $\frac{n}{g}$.    Let $E(gr)$ denote the number of equivalence classes of size $gr$ and the total number of equivalence classes by ${\cal E}$.  The values of $E(gr)$ and of ${\cal E}$ are given by (proof in the appendix):
\bean
E(gr) & = &   \frac{1}{gr}  \sum_{s\mid r}  \mu(s) {\frac{gr}{s} \choose \frac{(w+\gamma)gr}{ns} } \\
{\cal E} & = &  \sum_{r:  gr\mid n}  E(gr),
\eean
where $\mu(\cdot)$ denotes the M\"obius function.  Let  
\bean
\left\{ \pi^{(\ell)} \mid 1 \leq \ell \leq {\cal E} \right\} & = & \left\{ (\pi^{(\ell)}_1, \pi^{(\ell)}_2, \cdots, \pi^{(\ell)}_{w+\gamma} ) \mid  1 \leq \ell \leq {\cal E} \right\}, 
\eean
be the collection of subsets obtained by selecting one subset from each equivalence class. We will assume that the elements within each of the subsets $\pi^{(\ell)}$ are 
ordered in ascending numerical order, i.e., 
\bean
\pi^{(\ell)}_1 < \pi^{(\ell)}_2 < \cdots < \pi^{(\ell)}_{w+\gamma}, \ \  \text{  for all  } \ell. 
\eean
We will associate with each such subset, a collection of $n$ two-dimensional patterns, each having the same size $(w+\gamma)$.  This collection includes the fundamental pattern:
\bean
P^{(\ell)}(0) & = & \left\{ \left( i,\pi^{(\ell)}_i \right)  \mid 1 \leq i \leq w+\gamma \right\},
\eean
as well as its $n$ (columnar) cyclic shifts
\bean
P^{(\ell)}(\tau) & = &  \left\{ \left( i,\pi^{(\ell)}_i \oplus \tau \right)  \mid 1 \leq i \leq w+\gamma \right\}, \ \ 1 \leq \tau \leq (n-1) .
\eean
Given a pattern $P^{(\ell)}(\tau)$ we will refer to the $(w+\gamma)$-tuple
\bean
\pi^{(\ell)} \oplus \tau & = &  (\pi^{(\ell)}_1 \oplus \tau, \ \pi^{(\ell)}_2 \oplus \tau, \cdots, \ \pi^{(\ell)}_{w+\gamma} \oplus \tau ), 
\eean
as its (columnar) footprint.   Thus the footprint of a fundamental pattern $P^{(\ell)}(0)$ is simply given by $\pi^{(\ell)}$. 

\subsection{Layers of the Canonical Code}  \label{sec:can_gen_layers}

Let us define
\bean
L & = & \sum_{r: gr \mid n}  r E(gr) \\
V_1 & = & \text{lcm}(w,w+1,\cdots,w+\gamma-1)  \\
V & = &  \frac{V_1}{w} \\
M & = & L V .
\eean
Let us define the function $\{ \omega_{\ell} \mid 1 \leq \ell \leq {\cal E}\}$ by
\bean
\omega_{\ell} & = & r \text{ if the pattern $\pi^{(\ell)}$ has period $gr$} .
\eean
It follows that there are $E(gr)$ patterns corresponding to the value $\omega_{\ell}=r$.  Thus we can alternately express $L$ in the form
\bean
L & = & \sum_{r: gr \mid n}  r E(gr) \ = \ \sum_{\ell=1}^{\cal E} \omega_{\ell}.
\eean

Each code matrix $C$ is composed of $LV$ vertical stacked layers, each layer corresponding to a matrix $\{A^{(\ell,\omega,\nu)}\mid 1 \leq \ell \leq {\cal E} , \  1 \leq \omega \leq \omega , \ 1 \leq \nu \leq V\}$ of size $((w+\gamma) \times n)$.  Thus $C$ is of the form:
\bean
C & = & \left[   \begin{array}{c} 
A^{(1,1,1)} \\
\vdots \\
A^{(\ell,\omega,\nu)} \\
\vdots \\ 
A^{({\cal E},\omega_{\cal E},V)} \\
\end{array} \right] .
\eean
The entries of the $\{A^{(\ell,\omega,\nu)}\}$ are specified below. 


\subsection{Threading Codewords to Form a Layer} \label{sec:can_gen_threads}

The threading process is identical to the case of $(n, w+\gamma)=1$. We fix $(\ell,\nu)$ and the threading process in the $(\ell,\nu)$th layer is as follows.  Consider the collection of $n$ codewords $\{\underline{c}_{\tau}^{(\ell,\nu)} \}_{\tau=0}^{n-1}$ associated to a layer.  The symbols of the $\tau$th codeword $\underline{c}_{\tau}^{(\ell,\nu)}$, $0 \leq \tau \leq n-1$, are placed (in any arbitrary order) into the $n$ locations \bean
P^{(\ell)}(\tau) & = &  \left\{ \left( i,\pi^{(\ell)}_i \oplus \tau \right)  \mid 1 \leq i \leq w+\gamma \right\}, \ \ 1 \leq \tau \leq (n-1) ,
\eean
identified by the pattern $P^{(\ell)}(\tau)$.  The threading yields a $(w+\gamma \times n)$ matrix which we will denote by $A^{(\ell,\nu)}$.  We then repeat this process for each layer, i.e., for all pairs $(\ell,\nu)$.  It follows that a given pattern  $P^{(\ell)}(\tau)$ determines the ordering of code symbols in $\omega_{\ell}V$ layers.  In loose terms, each pattern is repeated $\omega_{\ell}V$ times and the parameters $\{\nu,\omega\}$ may hence be viewed as repetition parameters.   The repetition factor $\omega_{\ell}$ that is pattern dependent, will help as we shall see, ensure a larger code rate, whereas the constant repetition factor $V$ ensures a uniform download during node repair as in Section~\ref{sec:can_coprime}.  

Finally we vertically stack the matrices $A^{(\ell,\nu)}$ to obtain the code matrix. This completes specification of the code matrix $C$.   

\subsection{Parameters of the Canonical Code} \label{sec:can_gen_parameters}

\subsubsection{Parameters $n,\alpha$}  \label{sec:can_gen_nalpha}

Since the number of layers change, the parameter $\alpha$ is different from the case of $(n, w+\gamma)=1$. It is given by

\bean
\alpha & = & LV(w+\gamma) \\
& = & \frac{(w+\gamma) \cdot \text{lcm}(w,w+1,\cdots,w+\gamma-1)}{w}\sum_{r: gr \mid n}  r E(gr)   \\
& = & \frac{(w+\gamma) \cdot \text{lcm}(w,w+1,\cdots,w+\gamma-1)}{wg} \sum_{r: gr \mid n} \sum_{s\mid r}  \mu(s) {\frac{gr}{s} \choose \frac{(w+\gamma)gr}{ns} } 
\eean

\subsubsection{Determining Parameters $d$ and $\beta$} \label{sec:can_gen_dbeta}

Following the exact set of arguments in Sec.~\ref{sec:can_coprime_dbeta}, we can show that $d=n-\gamma$ and $\beta$ is given by 
\bea \label{eq:beta_small_d}
\beta & = & V (n,w+\gamma) \ \sum_{p=1}^\gamma {\gamma-1 \choose p-1} {n-\gamma-1 \choose w+\gamma-p-1} \frac{w}{w+\gamma-p}.
\eea

As a check, we note that each column contains $\alpha=LV (w+\gamma)$ symbols, each of which requires the transfer of $w$ symbols to enable repair.   Since there are a total of $(n-\gamma)$ helper nodes, we must have that 
\bean
\beta (n-\gamma) & = & w \alpha ,
\eean
i.e., $\beta$ must equal 
\bea \nonumber 
\beta & = & \frac{w}{(n-\gamma)} \alpha \\ \label{eq:beta_2_small_d} 
& = & \frac{w}{(n-\gamma)}  (w+\gamma)LV. 
\eea
It can be verified that the values for $\beta$ obtained in \eqref{eq:beta_small_d} and \eqref{eq:beta_2_small_d} are the same. 

\subsubsection{Determining $k, K$ and Code Rate $R$} \label{sec:can_gen_k}

Arguing as earilier, the scalar dimension of the code is clearly given by $K_c=LVnw$.  Not surprisingly, the rate $R$ of the code is given by $\frac{w}{w+\gamma}$.  

\section{Construction of $(n,k<d,d)$-Layered Regenerating Code} \label{sec:lay}

In this section, we will describe the construction of non-canonical layered regeneration code, \clrc for general parameter set $((n,k,d), (\alpha,\beta),K)$ regenerating codes, again indexed by two auxiliary parameters $w, \gamma$ satisfying $2 \leq w < k, 1 \leq \gamma \leq (n-k)$.

\subsection{Construction of the non-canonical code \clrc} \label{sec:lay_constrn}

\begin{figure}[h!]
\begin{center}
\includegraphics[width=6in]{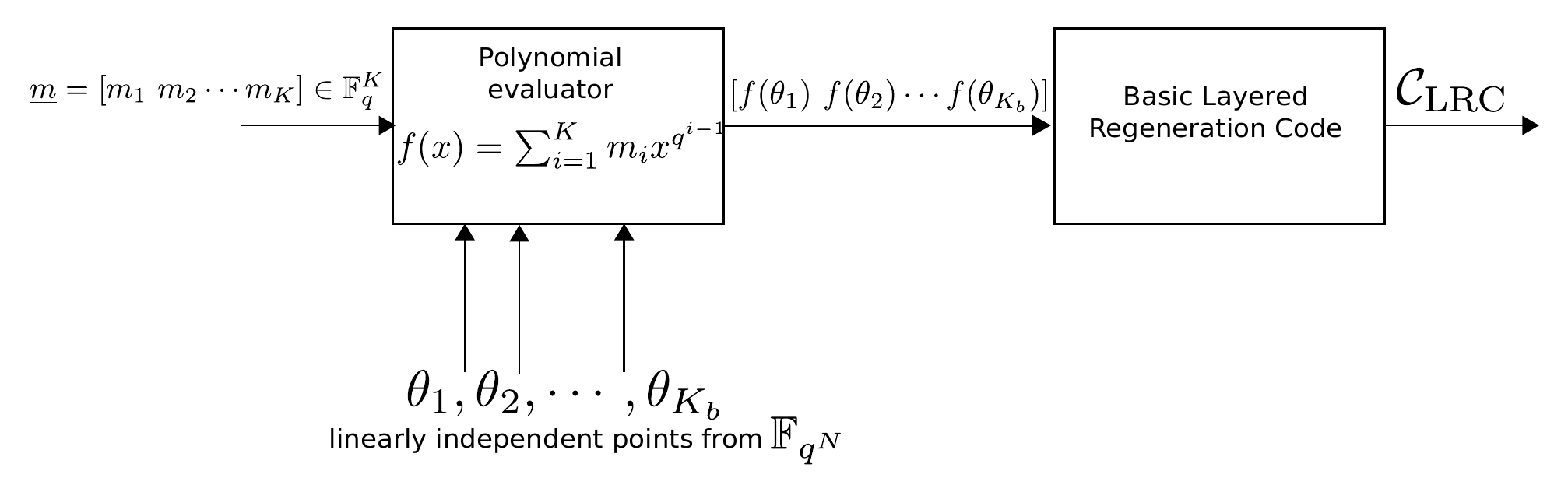}
\caption{Encoder of a Layered Regenerating Code.}
\label{fig:code_construction}
\end{center}
\end{figure}

The non-canonical regenerating code \clrc makes use of the canonical code code \ccan as shown in Fig.~\ref{fig:code_construction}.  It also makes use of linearized polynomials along the lines of their usage in \cite{RawKoySilVis}. Since the construction uses the canonical code, we need to consider the case of $(n,w+\gamma)=1$ and $(n,w+\gamma)>1$ separately. We will consider only the case of $(n,w+\gamma)=1$, and the general case follows accordingly. 

The $K$ message symbols $\{m_i\}_{i=1}^K$ of ${\cal C}_{\text{\tiny LRC}}$ are first used to construct a linearized polynomial 
\bean
f(x) & = & \sum_{i=1}^{K}m_i x^{q^{i-1}}.
\eean 
The linearized polynomial is then evaluated at $K_b$ elements $\{\theta_i\}_{i=1}^{K_b}$ of $\mathbb{F}_{q^N}$ which when viewed as vectors over $\mathbb{F}_q$, are linearly independent. The resulting $K_c$ evaluations $\{f(\theta_i)\}$ are than fed as input to an encoder for the canonical $\mathcal{C}$.  We set 
\bean
u_i & = & f(\theta_i), \  1 \leq i \leq K_c, \\
\underline{u} & = & (u_1, u_2, \ldots, u_{K_c}).
\eean

The non-canonical regenerating code is the output of the canonical code to the input $\underline{u}$.

\subsection{Parameters of \clrc} \label{sec:lay_parameters}

Clearly, the parameters $n, \alpha$ are exactly same as that of canonical code. First we proceed to relate $k$ and $K$ of \clrc. Towards that, we  being with presenting a generator-matrix view point of the canonical code. 

\subsubsection{Two generator matrices for the canonical code \ccan}  \label{sec:lay_genmatrix}

Thus far, we have described the code in terms of the structure of the codeword, viewed as a layered array.  Towards determining $k$ and $K$ of the code, we now turn to a generator matrix viewpoint of the code.  To obtain a generator matrix, one needs to vectorize the code matrix, thus replacing the code matrix by a vector of size $n \alpha=nLV(w+\gamma)$.  The generator matrix then describes the linear relation between the $LVw$ input symbols of the canonical code ${\cal C}$ and the $n\alpha$ output symbols.  Let us set $N_b  = n \alpha$ and recall that $K_b =  LVnw$.  Then the generator matrix is of size $(K_b \times N_b)$.  

The generator matrix is clearly dependent upon the manner in which vectorizing of the code matrix takes place.   We will present two vectorization and hence, two generator matrices:
\ben[(a)]
\item From the distributed storage network point of view, the natural vectorization is one in which the $N_b$ code symbols are ordered such that the first $\alpha$ symbols correspond to the elements of the first column vector (in top-to-bottom order), of the code matrix, the second $\alpha$ symbols correspond in order, to the elements of the second column vector etc.  Thus, under this vectorization, we will have that the first $\alpha$ columns of the generator matrix correspond to the first column vector of the code matrix and so on.  We will refer to this as the canonical vectorization of the code. In terms of the vector-code terminology introduced earlier, each set of columns of the generator matrix corresponding to a column of the code matrix, is referred to as a thick column of the generator matrix.  The code symbols associated to the $i$th thick column of the generator matrix are the code symbols stored in the $i$th storage node.  We will use $G$ to denote the generator matrix of the canonical code ${\cal C}$ under this vectorization.   
\item Next, consider a second vectorization of the canonical code ${\cal C}$ and hence, a different generator matrix.  The code symbols in the code matrix of the canonical code \ccan \ can be vectorized in such a manner that the resultant code vector is the serial concatenation of the $Mn$ codewords $\{\underline{c}_{\tau}^{(\ell,\nu)}\}$ of the code \cmds, each associated to a distinct message vector $\underline{u}_{\tau}^{(\ell,\nu)}$. Let $G_{\text{b-d}}$ denote the associated generator matrix of \ccan.  Clearly, $G_{\text{b-d}}$ has a block-diagonal structure:
\bea \label{eq:Gbd}
G_{\text{b-d}} & = & \left[
\begin{array}{cccc}
 \gmds &   &   & \\
  & \gmds  &   & \\
  &   &  \ddots  & \\
  & & & \gmds
\end{array}
\right].
\eea
Here \gmds\ denotes the generator matrix of the $[w+\gamma, w, \gamma+1]$-MDS code. It follows from this that the columns of $G_{\text{b-d}}$ associated to code symbols belonging to distinct MDS codewords are linearly independent.  Also, any collection of $w$ columns of $G_{\text{b-d}}$ associated with the same \cmds are linearly independent.  
\een
It is our intent to use the matrix $G$ for generating the canonical code \ccan and the matrix $G_{\text{b-d}}$ for analysis of \ccan.  We note that the two generator matrices $G$ and $G_{\text{b-d}}$ of the code \ccan \ differ only in the order in which the thin columns appear.

\subsubsection{Rank Accumulation in the Matrix $G$} \label{sec:lay_ra}

The matrix $G$ has the following uniform rank-accumulation property, namely that if one selects a set $S$ containing $s$ thick columns drawn from amongst the $n$ thick columns comprising $G$, then the rank the submatrix $G|_S$ of $G$ is independent of the choice of $S$. Hence the rank of $G|_S$ may be denoted as $\rho_s$, indicating that it just depends on the value of $s$. 

We now proceed to determine $\rho_s$. The value of $\rho_s$ depends on how the collection of thin columns in $S$ intersect with the blocks of $G_{\text{b-d}}$. For every thick column of $G|_S$, let us focus on a subset of thin columns corresponding to symbols from layers with a fixed value of $\nu$. We will refer to the submatrix of $G|_S$ thus obtained as $G^{(\nu)}|_S$. It is clear that rank of $G|_S$ is $V$ times the rank of $G^{(\nu)}|_S$. The intersection of $G^{(\nu)}|_S$ with blocks of $G_{\text{b-d}}$ can be sets of varying sizes, ranging from $0$ to $w+\gamma$. If the intersection is of size $p$, the rank accumulated is $\min\{p,w\}$, and thus it follows that 
\bea 
\label{eq:rho}
\rho_s & = & V \sum_{p=1}^{\min\{s,w+\gamma\}} {s \choose p} {n-s \choose w+\gamma-p} \min\{p,w\} . 
\eea

We define the rank-accumulation profile of the matrix $G$ as the collection of integers $\{a_i\}_{i=1}^{n}$ given by 
\bea
a_1 & = & \rho_1 \\
a_i & = & \rho_i-\rho_{i-1}, \ 2 \leq i \leq n. \label{eq:profile}
\eea
It is straightforward to see that 
\bean
a_i & = & \alpha, \ \ 1 \leq i \leq w, \\
a_i & = & 0, \ \ k+1 \leq i \leq n.
\eean
We will then have that 
\bean
\rho_s & = & \sum_{i=1}^s a_i, \ \ 1 \leq s \leq n.
\eean

\subsubsection{Parameters $k$ and $K$} \label{sec:lay_k}

Having described the rank accumulation profile of the canonical code, we are ready to relate $k$ and $K$ of the layered code \clrc. We begin with a useful lemma. 

\begin{lem} \label{lem:useful} 
Let $k_0$ be the smallest number of thick columns of the generator matrix $G$ of the canonical code \ccan\ such that the submatrix of $G$ obtained by selecting any $k_0$ thick columns of $G$ results in a submatrix of rank $\geq K$.  Then by connecting to any $k_0$ nodes associated to the regenerating code \clrc, a data collector will be able to recover the message symbols $\{m_i\}_{i=1}^{K}$.  
\end{lem}

\vspace*{0.2in}

\begin{proof}  Let $S$ be a collection of thick of $k_0$ thick columns of the matrix $G$ such that 
\bean
\text{Rank}\left( G|_S \right) & \geq & K.
\eean
The code symbols $(c_1,c_2,\cdots,c_n)$ of the layered regenerating code \clrc are related to $G$ as shown below
\bean
(c_1,c_2,\cdots,c_n) & = & [f(\theta_1) \ f(\theta_2) \ \cdots f(\theta_{K_b})] [G] .
\eean
Using linearity of $f(\cdot)$, we can write this as
\bean
(c_1,c_2,\cdots,c_n) & = & f(\underbrace{[\underline{x}_1 \ \underline{x}_2 \cdots \underline{x}_{K_b}]}_{(N \times K_b)}[G]) ,
\eean
in which $\underline{x}_i \in \mathbb{F}_q^N$ is the vector representation of the element $\theta_i \in \mathbb{F}_{q^N}$. Set
\bean
X & = & [\underline{x}_1 \ \underline{x}_2 \cdots \underline{x}_{K_b}].
\eean
Since the $\{ \underline{x}_i \}_{i=1}^{K_b}$ are linearly independent over $\mathbb{F}_q$, it follows that 
\bean
\text{Rank}\left( X \cdot G|_S \right) & = & \text{Rank}\left( G|_S \right) \\
& \geq & K.
\eean
Hence there are at least $K$ linearly independent columns in the matrix product $X \cdot G|_S$ and thus the computation $f \left( X \cdot G|_S \right)$ yields evaluations of $f(\cdot)$ in at least $K$ linearly independent points of $\mathbb{F}_{q^N}$. Since $f(\cdot)$ is of $q$-degree $(K-1)$, the coefficients of $f$ can be recovered from these $K$ evaluations. 
\end{proof}

\vspace*{0.2in}

It follows from the discussion above, that in order to relate the parameters $K,k$ of \clrc, it suffices to study the canonical code \ccan and determine the smallest number $k_0$ of columns of its generator matrix $G$, such that the corresponding sub matrix has rank at least $K$.  But from the uniform rank accumulation property of the generator matrix $G$ of the canonical code \ccan\, this is simply given by 
\bea \label{eq:value_k0}
k_0 & = & \min \left\{ k \mid \rho_k\ \geq \ K \right\}. 
\eea

Equivalently, the scalar dimension (or the filesize) of the layered regenerating code \clrc\ $K$ for a given value of $k$ is given by
\bea \label{eq:value_K}
K & = & V \sum_{p=1}^{\min\{k,w+\gamma\}}  \min\{w,p\}  {k \choose p}{n-k \choose w+\gamma-p} . 
\eea

\subsubsection{Parameters $d,\beta$} \label{sec:lay_dbeta}

From the discussion on rank accumulation profile, it follows that the scalar dimension $K$ will be strictly greater than $w\alpha$ when $w<k$, and hence we will have 
\beqn
w\alpha < K \leq k\alpha .
\eeqn
Thus, it is meaningful to have a scheme that repairs a failed node downloading $w\alpha$ symbols. Hence, we follow the same repair strategy as in the case of canonical code setting $d=n-\gamma$. We can repair any failed node downloading a fixed number $\beta$ of symbols from every helper node. The value of $\beta$ thus obtained would be
\bea 
\beta & = & V \ \sum_{p=1}^\gamma {\gamma-1 \choose p-1} {n-\gamma-1 \choose w+\gamma-p-1} \frac{w}{w+\gamma-p}.
\eea
It can also be checked that $(n-\gamma)\beta = w\alpha$.

\subsection{Some Remarks on the parameters of \clrc} \label{sec:lay_remarks}

The following remarks on parameters of \clrc \ are worth mentioning.

\begin{note} From the description in Sec.~\ref{sec:intro_reg}, it is clear that every regenerating code must satisfy
\bea \label{eq:alpha_range}
(d-k+1)\beta  & \leq  \ \alpha  \ \leq & d\beta .
\eea
Since layered regenerating codes have
\bean
d\beta & = & w\alpha \ ,
\eean
we must have 
\bean
k & \geq & 1 + d\left(\frac{w-1}{w} \right).
\eean
A lowerbound on $k$ imposes only a lower limit on the rate, and hence the above constraint does not come along with any penalty. 
\end{note}

\begin{note} In the construction, we have assumed the auxiliary parameter $w$ to be greater than $1$ because $w$ turns out to be the dimension of the common erasure code. Nevertheless, we can consider the extreme case of $w=1$, where the erasure code becomes a trivial repetition code. In addition, let us set $\gamma = 1$, and hence $w+\gamma=2$. Then for all odd valued $n$, $(n, w+\gamma) =1$ and hence in that case, we have
\bean
L & = & \frac{n-1}{2} , \\
V & = & 1 ,\\
\alpha & = & n-1, \\
d\beta & = & \alpha .
\eean
The code thus obtained is structurally similar to the repair-by-transfer MBR codes and differs only in that the underlying MDS code present in the construction of the repair-by-transfer MBR codes in \cite{ShaRasKumRam_rbt} is replaced here by an MDS code that is constructed using linearized polynomials. 
\end{note} 

\begin{note} If the linearized polynomial of $q$-degree $(K-1)$ used in the construction of \clrc \ is replaced by an (ordinary) polynomial of degree $(K-1)$, then one can then still go onto to obtain a regenerating code.  While this code will have smaller field size, it will however, have lesser rate in comparison to the code \clrc constructed here. 
\end{note}

\section{On the Optimality of the canonical code} \label{sec:opt}

In this section, we state two results pertaining to the performance of the canonical code against the storage-repair-bandwidth tradeoff. The first result shows that for any $(n,k,d=k)$ parameter set, we can construct canonical codes that performs better than what the space-sharing code achieves. In the second, we will establish the achievability of an interior point in the storage-repair-bandwidth tradeoff by an exact-repair code when $d=k=n-1$. The interior point we achieve is on the line-segment joining the MSR point and the next point of slope-discontinuity, where the non-achievability results established in \cite{ShaRasKumRam_rbt} does not apply. Both these results follow immediately from simple calculations.

\blem The $(n,k,d=k)$-canonical code operates at an $(\alpha, d\beta)$-point that lies between the MSR and MBR points, and performs better than the code that space-shares the MSR and MBR point, whenever $\gamma < w < k$. 
\elem
\bpf
For any regenerating code with $d=k$, we must have
\bean
\frac{\alpha}{d} \leq \beta \leq \alpha .
\eean
Since $w+\gamma \leq n$, we must have $w \leq n-\gamma = d = k$. Furthermore, $\beta = \frac{w}{d} \alpha$ for the canonical code. Thus it follows that code operate at a point between the MSR and MBR point.

From \cite{ShaRasKumRam_rbt}, we can express the space-sharing line in the form,
\bea
d\beta & = & \frac{ d(2K - k\alpha) }{k(d-k+1)}.
\eea
When $d=k$, it reduces to
\beqn
d\beta = 2K - k\alpha .
\eeqn
For the canonical code, we have 
\beqn
K = \left(\frac{w}{w+\gamma} \right) n\alpha ,
\eeqn
and hence, for it to perform better than the space-sharing code, we must have,
\beqn
w\alpha < 2 \left(\frac{w}{w+\gamma} \right) n\alpha - k \alpha .
\eeqn
It can be verified that the above condition holds whenever $(k-w)(w-\gamma) > 0$, which is true when $\gamma < w < k$.
\epf

\bcor When $n < 2k-1$, there exist exact-repair $(n,k,d=k)$-regenerating codes that operate between the MSR and the MBR point performing better than the space-sharing line.
\ecor
\bpf An integer value of $w$ satisfying $\gamma < w < k$ can be found when $n < 2k-1$. The statement follows from that.
\epf

\blem The $(n,n-1,n-1)$-canonical code achieves an interior point of the storage-repair-bandwidth tradeoff, that lies between the MSR point and the next point of slope-discontinuity specified by,
\bea
\label{eq:intpoint}
\alpha & = & (d-(k-2))\beta - \left(\frac{k-2}{k-1}\right)\beta.
\eea 
\elem
\bpf
The results in \cite{ShaRasKumRam_rbt} imply that the rank accumulation profile of a linear optimal regenerating code must satisfy,
\bean
a_p^{*} & = & \left\{ \begin{array}{ll} \min\{\alpha, (d-p+1)\beta\}, & 1 \leq p \leq k \\
					0 & k < p \leq n
					\end{array} \right. \\
	& = & \left\{ \begin{array}{ll} \alpha & 1 \leq p \leq \left\lfloor d - \left(\frac{\alpha}{\beta}\right) + 1 \right\rfloor \\
					(d-p+1)\beta & \left\lfloor d - \left(\frac{\alpha}{\beta}\right) +1 \right\rfloor < p \leq k \\
					0 & k < p \leq n 
					\end{array} \right. 
\eean
Thus a linear code is an optimal regenerating code if and only if it satisfies the above rank accumulation profile. For a regenerating code with $d\beta = w\alpha$, we calculate
\beq
a_p^{*}  =  \left\{ \begin{array}{ll} \alpha, & 1 \leq p \leq \left\lfloor d\left(\frac{w-1}{w}\right) +1 \right\rfloor \\
							   w\alpha - \frac{w(p-1)\alpha}{d}	& \left\lfloor d\left(\frac{w-1}{w}\right) +1 \right\rfloor < p \leq k 	\\
							   0 &  k < p \leq n 
							   \end{array} \right. \label{eq:opt}
\eeq

Now consider the $(n,n-1,n-1)$-canonical code. This means i.e., $\gamma=1$, and then it follows from \eqref{eq:profile} that the rank accumulation profile of the code, 
\beq
a_p  =  \left\{ \begin{array}{ll} \alpha, & 1\leq p \leq w \\
									\alpha - {p-1 \choose w}, & w+1 \leq p \leq k \\
									0. & k < p \leq n 
					\end{array} \right. \label{eq:can}
\eeq

For \eqref{eq:opt} and \eqref{eq:can} to match, it is also necessary to check that 
\bean
\left\lfloor d\left(\frac{w-1}{w}\right) +1 \right\rfloor =  w . \\
\Leftrightarrow \ \ \ d \in \{w, w+1\}
\eean

If we choose $w=d-1$, we obtain
\bean 
a_p = a_p^{*} = \alpha, \ \ 1 \leq p \leq w = k-1
\eean

Furthermore, we must check that $a_k = a_k^{*}$. 
\bea
a_k & = & \alpha - {k-1 \choose k-1} = \alpha -1 \label{eq:ak} ,\\
a_k^{*} & = & w\alpha - \frac{w^2\alpha}{w+1} \nonumber \\
& = & \frac{w\alpha}{w+1} . \label{eq:akopt}
\eea

For the canonical code, since $(n,w+1) = (w+2,w+1) = 1$,

\bean
\alpha & = & \frac{1}{n} {n \choose w+1} (w+1) \\
& = & \frac{1}{w+2} {w+2 \choose w+1} (w+1) \\
& = & w+1
\eean

Thus it follows that, 
\bean
a_k = a_k^{*} = w ,
\eean
showing that $(n,n-1,n-1)$-canonical code with $w=d-1$ is an optimal regenerating code. Since the accumulation profile has values $a_p = \alpha, 1 \leq p \leq (k-1)$ and $a_k = \alpha - 1$, it achieves a point between the MSR point and the next point of slope-discontinuity on the tradeoff. It can be calculated that the interior point thus achieved is specified by \eqref{eq:intpoint}.
\epf

\section{Codes with Canonical-Code-Locality} \label{sec:local}

In this section, we will briefly describe how it is possible to construct codes with locality in which each of the local codes is the canonical (layered regenerating)  code \ccan .  The same technique can also be used to generate codes with locality in which the local codes are the layered regeneration codes \clrc\!. 

\subsection{Locality in Vector Codes} \label{sec:locality_basics}

Let $\mathcal{C}$ be an $[n, K, d_{\text{min}}, \alpha]$ vector code over a field $\mathbb{F}_q$, possessing a $(K \times n\alpha)$ generator matrix $G$.  The $i^{\text{th}}$ code symbol, ${\bf c}_i$, is said to have (exact) $(r, \delta)$ locality, $\delta \geq 2$, if it is possible to puncture the code in coordinates corresponding to a set of indices $S$ with $i \in S$, such that the punctured code $\mathcal{C}|_S$ has length $r + \delta - 1$, and minimum distance $\delta$.
The code $\mathcal{C}$ is said to have $(r,\delta)$ all-symbol locality if all code symbols have $(r,\delta)$ locality.  The codes obtained through puncturing will be called local codes.  Our interest here is in the construction of a code with exact, all-symbol locality, whose local codes correspond to the canonical code \ccan\ introduced in Section~\ref{sec:can_coprime}. 

The property of locality allows to minimise the number of node accesses during node-repair. The concept of locality was introduced in \cite{GopHuaSimYek} for scalar codes for single erasures. Subsequently it was generalised to multiple erasures and later to vector codes; see \cite{PraKamLalKum}, \cite{PapDim}, \cite{KamPraLalKum} and \cite{RawKoySilVis}, \cite{OggDat}.  Codes combining benefits of regenerating codes and codes with locality are constructed in  \cite{RawKoySilVis}, \cite{KamPraLalKum} and ~\cite{KamPraLalKumSilRawKoyVis}.

\subsection{Code Construction} \label{sec:code_wth_locality}

Let $t\geq2$ and $\{\phi_i\}_{i=1}^{tK_c}$ a collection of elements in $\mathbb{F}_{q^N}$ and let $\{\underline{\phi}_i\}$ denote the representation of the $\{\phi_i\}$ as elements of $\mathbb{F}_q^N$.  Given a message vector $[m_1,m_2\ldots,m_K]^T$, we 
construct the linearized polynomial 
\bean
h(x) & = & \sum_{i=1}^{K} m_i x^{q^{i-1}}, \ \ m_i \in \mathbb{F}_{q^N}, \ \ N \geq t K_c,
\eean
and form the $tK_c$-tuple $[h(\phi_1),h(\phi_2),\ldots,h(\phi_{tK_c})]^T$.  This evaluation vector is then partitioned into $t$ evaluation vectors each counting $K_c$ components which are then fed to $t$ respective encoders for the canonical code.  The corresponding outputs of these encoders are then concatenated to form the desired codeword.   It can be shown that the resultant code is optimal in terms of having the best possible minimum distance for the given scalar dimension. 

\appendix

\section{Proof of Number of Orbits}

\section{Counting orbits of a particular size and total number of layers}

\blem  For $r$ such that $gr | n$, the number of equivalence classes of size $gr$ is given by 
\beqn
E(gr) =  \frac{1}{gr}  \sum_{s\mid r}  \mu(s) {\frac{gr}{s} \choose \frac{(w+\gamma)gr}{ns} } . 
\eeqn
In particular, the total number of equivalence classes ${\cal E}$ is given by
\beqn
{\cal E} =  \sum_{r\mid \frac{n}{g}}  E(gr). 
\eeqn
\elem

\bpf For $r$ such that $gr \mid n$, let $f_1(r)$ denote the number of equivalence classes of size less than or equal to $gr$.  Then $f_1(r)$ is given by 
\bean
f_1(r) & = & {gr \choose \frac{(w+\gamma)gr}{n} }.
\eean
Let $f_2(r)$ denote the number of patterns having size equal to $gr$.  
Then we have,
\beqn
f_1(r) = \sum_{s\mid r} f_2(s) 
\eeqn
and by M\"obius inversion, we obtain 
\beq
f_2(r) = \sum_{s\mid r} f_1\left(\frac{r}{s}\right) \mu(s), \label{eq:gt}
\eeq
where $\mu$ is the M\"obius function. Thus the number of equivalence classes of size $gr$ is given by
\bean
E(gr) & = & \frac{1}{gr} f_2(r) \\
& = & \frac{1}{gr} \sum_{s\mid r} f_1\left(\frac{r}{s}\right) \mu(s) \\
& = & \frac{1}{gr}  \sum_{s\mid r}  \mu(s) {\frac{gr}{s} \choose \frac{(w+\gamma)gr}{ns} } .
\eean
The result for the total number of equivalence classes follows immediately. 
\epf


\end{document}